\documentclass[reprint,aps,prb,english,superscriptaddress]{revtex4-2}

\usepackage{graphicx}
\usepackage{bm,amssymb,amsmath,mathrsfs,amstext,latexsym,physics}
\usepackage[T1]{fontenc}
\usepackage{float}
\usepackage[caption=false]{subfig}
\usepackage[usenames,dvipsnames]{xcolor}
\usepackage[colorlinks=true,citecolor=blue,urlcolor=blue,linkcolor=blue]{hyperref}
\usepackage[normalem]{ulem}
\usepackage{multirow}
\usepackage{makecell}

\usepackage{scalerel}
\DeclareMathOperator*{\Bigcdot}{\scalerel*{\cdot}{\bigodot}}

\usepackage{bm, physics}

\newcommand{\fn}[2]{\mathinner{#1\mathopen{\left(#2\right)}}}
\newcommand{\vect}[1]{{\bf #1}}
\newcommand{\E}[1]{\left\langle#1\right\rangle}
\newcommand{\weighted}[3]{\fn{#1_{#2}}{#3}}
\newcommand{\field}[2]{\weighted{n}{#1}{#2}} 
\newcommand{\CHI}[2]{\fn{{\chi}_{#1}}{#2}}
\newcommand{\Z}{\mathbb{Z}}

\usepackage{soul}
\setstcolor{red}
\setul{0}{0.3ex}

\begin{document}

\title{Hyperuniformity of Weighted Particle Systems
}

\author{Salvatore Torquato}
\email[Corresponding author: ]{torquato@electron.princeton.edu}
\affiliation{Department of Chemistry, Princeton University, Princeton, New Jersey 08544, USA}
\affiliation{Department of Physics, Princeton University, Princeton, New Jersey 08544, USA}
\affiliation{Princeton Materials Institute, and Program in Applied and Computational Mathematics, Princeton University, Princeton, New Jersey 08544, USA}

\author{Jaeuk Kim}
\affiliation{Department of Chemistry, Princeton University, Princeton, New Jersey 08544, USA}
\affiliation{Department of Chemical and Biomolecular Engineering, Korea Advanced Institute of Science and Technology (KAIST), Daejeon 34141, Republic of Korea}
\affiliation{GIST InnoCOREAI-Nano Convergence Institute for Early Detection of Neurodegenerative Diseases, Gwangju Institute of Science and Technology, 61005 Gwangju, Republic of Korea}

\author{Michael A. Klatt}
\affiliation{German Aerospace Center (DLR), Institute for AI Safety and Security, Wilhelm-Runge-Strasse 10,
89081 Ulm, Germany}
\affiliation{German Aerospace Center (DLR), Institute of Frontier Materials on Earth and in Space, Functional, Granular, and Composite Materials, 51170 Cologne, Germany}
\affiliation{Department of Physics, Ludwig-Maximilians-Universität München, Schellingstrasse 4, 80799 Munich, Germany}

\author{Roberto Car}
\affiliation{Department of Chemistry, Princeton University, Princeton, New Jersey 08544, USA}
\affiliation{Princeton Materials Institute, and Program in Applied and Computational Mathematics, Princeton University, Princeton, New Jersey 08544, USA}

\author{Paul J. Steinhardt}
\affiliation{Department of Physics, Princeton University, Princeton, New Jersey 08544, USA}

\date{\today}

\begin{abstract}

Hyperuniform particle arrangements are characterized by a local number variance within a spherical window of radius $R$ that grows more slowly than the volume of the window, i.e.,  $R^{d}$,  in $d$-dimensional Euclidean space. We generalize this concept to describe the large-scale behavior of particle systems in which particles carry weights: internal degrees of freedom such as scalars (charges and masses), vectors (electric dipole moments,  velocities,
and torques), pseudovectors (spins and angular momenta), directors  (bond orientations), tensors (quadrupole moments) or extrinsic local attributes (Voronoi-cell characteristics). The underlying hyperuniform arrangement may be ordered (crystals, quasicrystals) or disordered, the latter of which has been extensively studied for its novel properties. Our generalization extends hyperuniformity from fluctuations in particle positions to fluctuations in the spatial distribution of weights, and examines how weighted fluctuations compare to their unweighted counterparts. We derive generalized weighted pair correlation functions, autocovariance functions, and spectral functions, and show their relation to formulas for the local variance in weighted many-particle systems.  Then we  apply
our formalism to determine the hyperuniformity/nonhyperuniformity of bond-orientational ordered
phases, dipolar liquid water, Voronoi-cell volumes and certain ionic liquids in various Euclidean
space dimensions. We demonstrate that hyperuniformity in the particle system does not necessarily translate to hyperuniformity of the weighted system. In fact, cases exist where a hyperuniform particle system becomes antihyperuniform when weighted, and others where nonhyperuniform or antihyperuniform particle systems yield hyperuniform weighted systems. This theoretical framework provides a road map for quantifying large-scale fluctuations in weighted many-particle systems, offering a powerful tool for identifying systems with novel physical properties.

\end{abstract}

\maketitle

\section{Introduction}
Hyperuniform particle systems in $d$-dimensional Euclidean space $\mathbb{R}^d$ are characterized by an anomalous suppression
of large-scale particle density fluctuations compared to typical disordered systems ({\it i.e.,} ordinary liquids and structural glasses) \cite{To03a,To18a}. More precisely, 
the local particle number variance within a spherical window of volume $R^d$, $\sigma_N^2(R) \equiv \langle N(R)^2 \rangle - \langle N(R) \rangle^2$, has the property that $\sigma_N^2(R)/R^d \rightarrow 0$ as $R\rightarrow \infty$.
Hyperuniform systems include all perfect crystals and many quasicrystals 
as well as a spectrum of disordered states in which the disorder is correlated on large scales.
Hyperuniformity can be equivalently stated as the vanishing of the structure factor $S(\mathbf{k})$ as the wave number $k=|\mathbf{k}|$ goes to zero, i.e., $\lim_{|\mathbf{k}| \to 0} S(\mathbf{k}) = 0$.

Over the past two decades, disordered hyperuniformity has been discovered in a variety of different
contexts across the physical, mathematical and biological sciences, 
including maximally random jammed states \cite{Do05c,Ma23}, classical disordered ground states \cite{To15}, random matrices \cite{To08b},
number theory \cite{To19}, critical absorbing states \cite{He15}, active particle systems \cite{Le19a,Ba24}, torque fluctuations in frictional matter \cite{Sha25}, biological systems \cite{Ji14,Ma15,Hu21}, and ecosystems \cite{Ge23}. Disordered hyperuniform 
systems have also attracted great attention in recent years because they are endowed with novel optical, acoustic, transport, mechanical
and chemical properties \cite{Fl09b,Le16,Fr17,Xu17,Gk17,To18c,Ro19,Ki20a,Ch21,Al23,Ki23}.

\begin{figure}[ht]
\centerline{\includegraphics[width=0.95\linewidth]{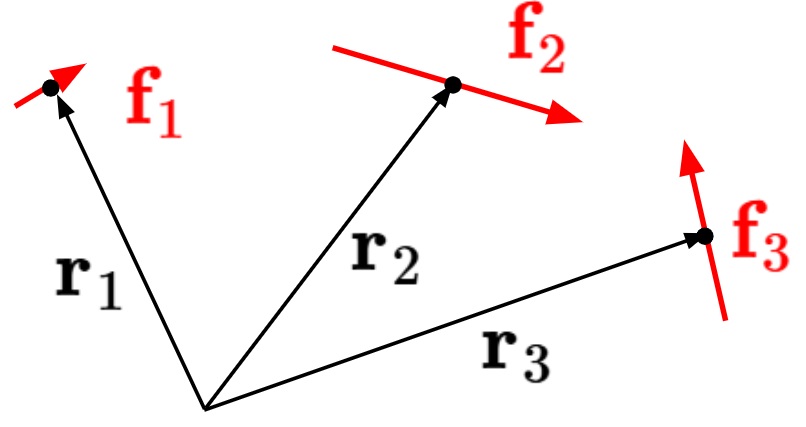}}
\caption{ Schematic of a point configuration with (complex-valued) vector weights of different magnitudes and directions.
}
\label{vector-weights}
\end{figure}

In this paper, we generalize the hyperuniformity concept to weighted particle systems in which
we derive an analogous theoretical formalism to characterize the large-scale spatial fluctuations in the weights. The weights may be  internal degrees of freedom assigned to each particle that can take the form of scalars, vectors, pseudovectors, directors, or tensors.
Examples include real $d$-dimensional vectors (see Fig. \ref{vector-weights}), such as dipoles \cite{Ki39,Gr97,Be93b,Sh07}, torques in frictional granular packings \cite{Sha25}, spin vectors \cite{Ud21,Me84,Le19c}, bond vectors in polymer chains \cite{Fl69,Se10}, orientation vectors of anisotropic particles (liquid crystals \cite{St74,De95}) or velocity vectors in active particle systems \cite{Mar13}, fluidized beds, sedimenting suspensions, and granular flows \cite{Lad96,Ha15,Cow00,Yu20}.  The weights may be a real scalar, which we denote by $f_i$, such
as those in charged-particle systems \cite{Leb83,An12b}, point-particle masses \cite{Go08}, excess contact numbers of particles in jammed packings \cite{He18b,Ik20,Sh25b}, or the volume or surface area of a Voronoi cell associated with the $i$th point \cite{Kla14,Fa14,Ch21c}.
An example of a complex scalar weight associated with a point is $\exp(im\theta)$, which has been used to characterize $m$-fold 
bond-orientational order
in two dimensions \cite{Ha78,Lev86}. This partial list of possible types of weights is summarized in Table \ref{tab:examples}.

\begingroup 
\begin{table*}[ht]
\renewcommand{\arraystretch}{1.5}
\caption{Examples of weighted point (particle) configurations.
\label{tab:examples}}
\begin{tabular}{c|c|c}
\hline
Examples & \multicolumn{2}{c}{Weight types}\\
\hline 
\makecell{
Charged systems (e.g., Coulombic systems and ionic liquids)
\cite{Ma80,Leb83,Le97,An12b,Le19c} \\
Point-particle masses (e.g., classical mechanics of celestial bodies) \cite{Go08} \\ 
 Particle cell statistics (e.g., Minkowski functionals of tessellations) \cite{Sc11,Fa14,Kla14,Al17,Ch21c} \\
Excess contact numbers in jammed packings \cite{He18b,Ik20, Sh25b}
 } 
& Real-valued & \multirow{2}{*}{Scalar}  
\\
\cline{1-2}
\makecell{
Bond-orientationally ordered phases (e.g., hexatic, nematic, etc.)
\cite{Ha78,Lev86} \\
Quasicrystals \cite{Lev86} \\
} & Complex-valued &
\\
\hline
\makecell{
Dipolar systems (e.g., water and ferrofluids) \cite{Ki39,Gr97,Be93b,Sh07}  \\
Orientation vectors in liquid crystals \cite{St74,De95}  \\
Bond vectors in polymer chains \cite{Fl69,Se10}  \\
Velocity vectors in active particle systems \cite{Mar13}  \\
Velocity vectors in fluidized beds, sedimenting suspensions and granular flows \cite{Lad96,Ha15,Cow00} \\
Torques in frictional granular packings \cite{Sha25}\\
Spin systems (e.g., spin ices and spin glasses) \cite{Pa45,Ud21,Me84,Le19c} }

& Real-valued &\multirow{2}{*}{Vector}
\\
\hline
\end{tabular}
\end{table*}
\endgroup

We introduce basic definitions in Sec. \ref{sec:definitions} and 
 derive the appropriate generalized
pair correlation functions, autocovariance functions and corresponding spectral functions in Sec. \ref{auto-spec}. Subsequently, in Sec. \ref{variance}, we show how these pair statistics are linked to the appropriately generalized formulas for the local variance of such weighted point configurations.
All of these derivations follow closely the original formulation of hyperuniformity for unweighted point configurations
\cite{To03a} as well as its generalizations to two-phase media, scalar fields and vector fields \cite{Za09,To16a}.

We then apply the theoretical formalism to ascertain
the hyperuniformity or nonhyperuniformity of bond-orientationally ordered phases, dipolar liquid water, Voronoi-cell volumes and certain ionic liquids
in various Euclidean space dimensions in Secs. \ref{revisit}, \ref{volume} and \ref{charges}. Among other objectives, we quantitatively probe how the fluctuations imposed by the weights differ from their unweighted counterparts. This is done by quantifying the differences in the corresponding spectral functions, 
including the changes in their scaling exponents in the zero-wave-number limit, and the differences
in the corresponding local variances.

We begin by revisiting previous well-established
results for bond-orientationally ordered (hexatic/nematic/tetratic) phases in two dimensions (as predicted by the Kosterlitz-Thouless-Halperin-Nelson-Young (KTHNY) theory \cite{Ko73,Ha78,Ne79}) and dipoles in water and analyze them under the hyperuniformity lens of weighted point configurations (see Sec. \ref{revisit}). While the unweighted particles in the bond-orientationally ordered phases
are nonhyperuniform (fluctuations grow like the window volume), they are antihyperuniform (weighted fluctuations that grow faster than the window volume) with respect to twofold, fourfold and sixfold bond-orientational order.

 Subsequently, we carry out additional calculations to quantify
the weighted total correlation function $h_{\psi_6}({\bf r})$ associated with sixfold bond-orientational order 
in the solid phase of 2D hard disks as well as certain 2D hyperuniform packings. 
While the unweighted particles in the 2D solid phase
are nonhyperuniform, they are antihyperuniform for particles weighted by the sixfold bond-orientational order parameter.

We also apply our analysis to the case of dipolar water. 
The unusually high dielectric constant of water is due to the strong correlations in the orientations 
of dipole moments of the water molecules that is imposed by the hydrogen-bond network \cite{Pa45}. In Sec. \ref{revisit}, we  reexamine
simulation data for liquid water presented in Ref. \cite{Sh07}
using our present theoretical formulation. We conclude that
liquid water in equilibrium is typically nonhyperuniform
with respect to dipole moments, as is its unweighted counterpart.

In Sec. \ref{volume}, we consider Voronoi tessellations of statistically homogeneous point configurations in $\mathbb{R}^d$ and assign to each particle a scalar weight equal to its corresponding Voronoi-cell volume.  We  present  
theoretical and numerical results for the weight-averaged autocovariances, spectral densities and local variances for certain 1D, 2D and 3D models of hyperuniform, standard nonhyperuniform and antihyperuniform unweighted point configurations. These models include 
Poisson point configurations, random sequential addition packings, hyperplane intersection process, stealthy hyperuniform point configurations, and maximally random jammed states. We 
show that in each case the weighted point configurations are strongly hyperuniform in 
that the variance grows like the surface area  (perimeter in two dimensions) of the window.

In Sec. \ref{charges}, we introduce the concept of scalar weights derived from the ``excess'' number of sides of Voronoi cells associated with 2D statistically homogeneous disordered point configurations and how they can be mapped to certain charged 2D systems with effective long-range interactions.
The excess side (edge) number of such a cell  is the difference between 
the number of sides  and the mean of six, which is set by Euler's formula for disordered tessellations in which no four points are cocircular and no three points are collinear. We analyze the same models as in Sec. \ref{volume}
and show that they are hyperuniform such
that their weighted variances grow like the surface area   of the window
with respect to charges  even if the unweighted configurations are antihyperuniform, nonhyperuniform or hyperuniform.  We briefly describe an extension of
our procedure to {\it weighted random fields} in Appendix F.

\section{Basic Definitions}
\label{sec:definitions}

Consider an ensemble of $N$ particles whose centroids are located at positions ${\bf r}^N \equiv {\bf r}_1,{\bf r}_2,\ldots, {\bf r}_N$ 
with vector weights ${\bf f}^N \equiv {\bf f}_1,{\bf f}_2,\ldots, {\bf f}_N$, respectively, in a region $\cal V$ of volume $V$ in $d$-dimensional Euclidean space $\mathbb{R}^d$ \footnote{Mathematically, we refer to them as point configurations, which are also known as point processes.}.
The weight $\bf f_i$ associated with the $i$th point could represent a tensor of rank $n$ with complex matrix entries, but for simplicity and concreteness,
we take it to be a complex $d$-dimensional vector.
In the mathematics literature, weighted point configurations are known
as {\it marked point processes} and have received considerable attention \cite{Chi13,Bj24}.


Let ${\bf Q}_i$ represent the generalized $2d$-dimensional vector coordinate of the $i$th particle that describes both
the position vector ${\bf r}_i$ and vector weight ${\bf f}_i$, and $\dd{\bf Q}_i$ denote the corresponding $2d$-dimensional volume element.
The weighted point configuration is statistically characterized by the {\it specific} probability density $P_N({\bf Q}^N)$,
where 
$P_N({\bf Q}^N) \; \dd{\bf Q}^N$ gives the probability of finding
the center of particle 1 in the volume element
$\dd{\bf Q}_1$ about ${\bf Q}_1$, 
the center of particle 2 in the volume element
$\dd{\bf Q}_2$ about ${\bf Q}_2$,
\ldots, and the center of particle $N$
in volume element $\dd{\bf Q}_N$ about ${\bf Q}_N$, and
$\dd{\bf Q}^N\equiv \dd{\bf Q}_{1}\dd{\bf Q}_2\cdots \dd{\bf Q}_N$.
Since $P_N$ is a probability density function, it normalizes to unity, i.e.,
\begin{equation}
\int P_N({\bf Q}^N) \;\dd{\bf Q}^N = 1.
\label{normal1}
\end{equation}
The ensemble average of a function $H({\bf Q}^N)$ is given by
\begin{equation}
\langle H({\bf Q}^N) \rangle = \int \cdots \int H({\bf Q}^N) P_N({\bf Q}^N) \dd{\bf Q}^N.
\end{equation}
The reduced probability density function for $n <N$ points
is obtained in the usual way:
\begin{equation}
P_n({\bf Q}^n)=\int \cdots \int P_N({\bf Q}^N) \;\dd{\bf Q}_{n+1} \cdots \dd{\bf Q}_N
\end{equation}
The reduced {\it generic} 
$n$-particle probability density function $\rho_n({\bf Q}^n)$ is proportional to the
probability density function associated with finding 
any $n$ points with configuration ${\bf Q}^n$:
\begin{equation}
\rho_n({\bf Q}^n) = \frac{N!}{(N-n)!} P_n({\bf Q}^n)
\end{equation}
and obeys the normalization condition
\begin{equation}
\int \cdots \int
\rho_{n}({\bf Q}^n)\dd{\bf Q}^n=\frac{N!}{(N-n)!},
\label{norm1}
\end{equation}
where we call $\rho_{n}({\bf Q}^n)$ \textit{the generalized $n$-particle probability density function} for weighted point configurations.

\smallskip

For   statistically homogeneous systems,
$\rho_{n}({\bf Q}^n)$
is translationally invariant with respect to positions ${\bf r}^n$ and, hence, depends only on the relative displacements
of the positions with respect to some chosen origin, say ${\bf r}_1$.
In particular, the one-particle 
function $\rho_1({\bf Q}_1)$ is independent of position ${\bf r}_1$, and, hence,
\begin{align}
N =\int \rho_1({\bf Q}_1) \dd{\bf Q}_1 
= V \int \rho_1({\bf f}_1) \dd{\bf f}_1 .
\end{align}
This allows one to define the singlet probability density function
for the weights to be
\begin{equation}
p({\bf f}) =  \frac{1}{N} \rho_1({\bf Q})=\frac{V}{N} \rho_1({\bf f}),
\end{equation}
which must normalize to unity, as required.
Therefore, in the thermodynamic limit
\begin{equation}
\rho_1({\bf Q}_1) = \rho\; p({\bf f}_1),
\label{thermolimit}
\end{equation}
where
\begin{equation}
\rho \equiv \lim_{ N,V\rightarrow\infty} \frac{N}{V} 
\end{equation}
is the number density.
Similarly, the two-particle probability density
function depends on the displacement vector ${\bf r}_{12} \equiv {\bf r}_2 -{\bf r}_1$ as well as the two weights (i.e., $\vect{f}_1$ and $\vect{f}_2$),
and, hence, we write $\rho_{2}({\bf Q}_1, {\bf Q}_2)= \rho_{2}({\bf r}_{12};{\bf f}_1,{\bf f}_2)$.
Analogously, we define \textit{the generalized pair correlation function} as follows:
\begin{align} \label{eq:generalized-g2}
g_{2,\vect{f}}(\vect{r};\vect{f}_1,\vect{f}_2) \equiv \frac{\rho_{2}({\bf r}_{12};{\bf f}_1,{\bf f}_2)}{\rho^2}.
\end{align}
Note that, when there is no long-range order, $g_{2,\vect{f}}({\bf r};{\bf f}_1,{\bf f}_2)$ tends to $
p({\bf f}_1)p({\bf f}_2)$ in the limit $|
{\bf r}| \to \infty$. 
Using this pair correlation function, the {\it weight-averaged pair correlation function} $\fn{g_{2,\vect{f}}}{\vect{r}}$ is defined to be
\begin{align}  \label{eq:g_2f}
\fn{g_{2,\vect{f}}}{\vect{r}} \equiv
\iint ({\bf f}_1^*\Bigcdot{\vect{f}}_2) \fn{g_{2,\vect{f}}}{\vect{r};\vect{f}_1,\vect{f}_2} \dd{\vect{f}_1} \dd{\vect{f}_2},  
\end{align}
where $\Bigcdot$ denotes an inner product of two vectors, meaning that this function has the dimensions of the square of the weights.
We see that, when there is no long-range order,
$\fn{g_{2,\vect{f}}}{\vect{r}}$ tends to $ |\expval{\bf f}| ^2$ in the limit $|{\bf r}| \to \infty$.
Note that for unweighted situations, relation (\ref{eq:g_2f}) reduces to $g_2({\bf r})\equiv \rho_2({\bf r})/\rho^2$, which is the standard definition for
point configurations without weights \cite{Han13}.

\section{Autocovariance Function and Spectral Density}
\label{auto-spec}

Consider the {\it weighted} Dirac measure at position ${\bf x}$:
\begin{equation}
\field{\vect{f}}{\vect{x}} = \sum_{j = 1}^N {\vect{f}}_j \fn{\delta}{\vect{x}-\vect{r}_j},
\label{n}
\end{equation}
which is a $d$-dimensional vector.
From Campbell's theorem \cite{Chi13}, the ensemble average of $\field{\vect{f}}{\vect{x}}$ for a statistically homogeneous system is given by
\begin{align}
\E{\field{\vect{f}}{\vect{x}}} 
= & 
\sum_{j = 1}^N \int \vect{f}_j \fn{\delta}{\vect{x} - \vect{r}_j} 
\fn{P_N}{\vect{Q}^N} \dd{\vect Q}^N \nonumber \\
=&
\sum_{j = 1}^N \int \vect{f}_j
\fn{P_1}{\vect{x};\vect{f}_j}\, \dd{\vect{ f}}_j
\nonumber \\
=& 
\int {\vect f}_1\, \fn{\rho_1}{\vect{x};\vect{f}_1} \dd{\vect{f}_1} 
=
\rho \int {\bf f}\, \fn{p}{\vect{f}} \dd{\vect{f}} 
\nonumber \\
=& 
\rho \E{\vect f},
\end{align}
where $ \E{\vect f}$ is the ensemble average of the vector weight $\bf f$.
Here, we have used Eq. (\ref{thermolimit}) and have taken the thermodynamic limit.
Note that $\E{\vect f}$ for an {\it ergodic} system is also given by the following arithmetic
average for a single realization in the infinite-$N$ limit:
\begin{equation}
 \E{\vect f} = \lim_{N \to \infty} \frac{1}{N} \sum_{j=1}^N {\bf f}_j.
\end{equation}

\begin{widetext}
Now consider the following {\it weight-averaged autocovariance function} for statistically homogeneous systems:
\begin{align}
\CHI{\vect{f}}{\vect{r}} &\equiv \E{\qty[\field{\vect{f}}{\vect{x}_1} - \E{\field{\vect{f}}{\vect{x}_1}}]^* \Bigcdot \qty[\field{\vect{f}}{\vect{x}_2} - \E{\field{\vect{f}}{\vect{x}_2}}]},
\label{autoco}
\end{align}
where ${\bf r}={\bf x}_2-{\bf x}_1$, implying that
$\CHI{\vect{f}}{\vect{r}}$ is a scalar function. We could further generalize the autocovariance function to be a dyadic product
of the two random vector variables appearing in relation (\ref{autoco}), but for simplicity, we focus here on their inner product and, hence, a scalar autocovariance function.
Using relation \eqref{n}, we obtain
\begin{align}
 \CHI{\vect{f}}{\vect{r}} 
 =&
 \E{\sum_{j = 1}^N |{\vect f}_j|^2 \,\fn{\delta}{\vect{x}_1 - \vect{r}_j}\, \fn{\delta}{\vect{x}_2 - \vect{r}_j}}
 +
 \E{\sum_{j \neq k}^N ({\bf f}_j^*\Bigcdot{\bf f}_k) \,\delta({\bf x}_1 - {\bf r}_j)\, \delta({\bf x}_2 - {\bf r}_k)}
 -\rho^2 \E{{\bf f}^*} \Bigcdot \E{\vect{f}}
 \nonumber \\
 =&
 \sum_{j = 1}^N \int |{\bf f}_j|^2 \,\delta({\bf x}_1 - {\bf r}_j)\, \delta({\bf x}_2 - {\bf r}_j) P_1({\bf Q}_j) \dd{\bf Q}_j 
 + 
 \sum_{j \neq k}^N \int \int ({\bf f}_j^*\Bigcdot{\bf f}_k) \,\delta({\bf x}_1 - {\bf r}_j)\, \delta({\bf x}_2 - {\bf r}_k) P_2({\bf Q}_j,{\bf Q}_k) \dd{{\bf Q}_j ^*} \dd{\bf Q}_k 
 - \abs{\rho\E{\vect{f}}}^2
 \nonumber \\
 =& 
 \fn{\delta}{\vect{r}}
 \int |{\vect{f}}|^2 \, \fn{\rho_1}{\vect{x};\vect{f}} \dd{\bf f}
 + 
 \int \int ({\bf f}_1^*\Bigcdot{\vect{f}}_2) \, \fn{\rho_2}{\vect{r};{\bf f}_1,\vect{f}_2} \dd{\bf f}_1 \dd{\vect{f}_2} 
 - \abs{\rho\E{\vect{f}}}^2 
 \nonumber \\
 =& 
 \rho \fn{\delta}{\vect{r}}
 \int |{\bf f}|^2 \, p({\bf f}) \dd{\bf f}
 + 
 \int \int ({\bf f}_1^*\Bigcdot{\vect{f}}_2) \, \fn{\rho_2}{\vect{r};\vect{f}_1,\vect{f}_2} \dd{\vect{f}_1} \dd{\vect{f}_2} 
 - \abs{\rho\E{\vect{f}}}^2 
 \nonumber \\
 =& 
 \rho \E{\abs{\vect{f}}^2} \fn{\delta}{\vect{r}}
 + 
 \int \int ({\bf f}_1^*\Bigcdot{\vect{f}}_2) \, \fn{\rho_2}{\vect{r};\vect{f}_1,\vect{f}_2} \dd{\vect{f}_1} \dd{\vect{f}_2} 
 - \abs{\rho\E{\vect{f}}}^2 
 \nonumber\\
 =& 
 \rho \E{\abs{\vect{f}}^2} \fn{\delta}{\vect{r}}
 + 
 \int \int ({\bf f}_1^*\Bigcdot{\vect{f}}_2) \, \qty [\fn{\rho_2}{\vect{r};\vect{f}_1,\vect{f}_2} - \fn{\rho_1}{\vect{r}_1;\vect{f}_1} \fn{\rho_1}{\vect{r}_2;\vect{f}_2}] \dd{\vect{f}_1} \dd{\vect{f}_2} .
\label{chi-r-1}
\end{align}

\end{widetext}

\begingroup 
\begin{table*}[th!]
\renewcommand{\arraystretch}{1.5}
\caption{Summary of the definitions of the weight-averaged correlation functions in direct space and their extreme behaviors. The values of the functions in the limit $r \to \infty$ assume
no long-range order.
\label{tab:direct} }
 \begin{tabular}{c|c c c c c}
 \hline 
 Weight-averaged functions& Dimensions& $r=0$& $r\to \infty$& \makecell{Unweighted counterpart} \\
 \hline 
 \makecell{
 Autocovariance, \\
 $\fn{\chi_{\vect{f}}}{r} \equiv \expval{[n_{\vect{f}}(\vect{r}+\vect{r}') - \expval{n_{\vect{f}}(\vect{r}+\vect{r}')} ]^*\cdot [n_{\vect{f}}(\vect{r}') - \expval{n_{\vect{f}}(\vect{r}')} ]}$}
 &
 $\qty[ \frac{\text{weight}}{\text{volume}}]^2$
 &
 $\rho \expval{\abs{\vect{f}}^2} \delta(\vect{r})$
 &
 0
 &
 $\fn{\chi}{\vect{r}}$
 \\

 \hline 
 \makecell{
 Pair correlation,
 \\
 $\fn{g_{2,\vect{f}}}{r} \equiv \iint \vect{f}_1^* \cdot \vect{f}_2 \rho_2(\vect{r};\vect{f}_1,\vect{f}_2) \dd{\vect{f}_1}\dd{\vect{f}_2} /\rho^2$
 \\
 ${\color{white}g_{2,\vect{f}}(\vect{r})}~ = \chi_{\vect{f}}(\vect{r})/\rho^2 - \expval{\abs{\vect{f}}^2} \delta(\vect{r})/\rho + \abs{\expval{\vect{f}}}^2$
 }
 &
 $\qty[ \text{weight}]^2 $
 &
 $0$
 &
 $\abs{\expval{\vect{f}}}^2$
 &
 $g_2(\vect{r})$
 \\  

 \hline 
 \makecell{
 Total correlation,
 \\
 $\fn{h_{\vect{f}}}{\vect{r}} \equiv g_{2,\vect{f}}(\vect{r}) - \abs{\expval{\vect{f}}}^2$
 \\
 $~~~~~~~~~~~{\color{white}\fn{h_{\vect{f}}}{r}}~ = \chi_{\vect{f}}(\vect{r})/\rho^2 - \expval{\abs{\vect{f}}^2} \delta(\vect{r})/\rho $
 }
 &
 $\qty[ \text{weight}]^2$
 &
 $-\abs{\expval{\vect{f}}}^2$
 &
 $0$
 &
 $h(\vect{r})$
 \\   
 \hline 

 \end{tabular}

\end{table*}
\endgroup

\begingroup 
\begin{table*}[bt]
\renewcommand{\arraystretch}{1.5}
 \caption{Summary of the definitions of the weight-averaged spectral functions in Fourier space and their behaviors
at extreme values. 
\label{tab:Fourier}}
 \begin{tabular}{c|c c c c c}
 \hline 
 Weight-averaged functions& Dimensions& $k=0$& $k \to \infty$& \makecell{Corresponding unweighted quantity} \\
 \hline 
 \makecell{
 Spectral density, \\
 $\fn{\tilde{\chi}_{\vect{f}}}{\vect{k}} \equiv \int \fn{\chi_{\vect{f}}}{\vect{r}} \exp(-i\vect{k}\cdot\vect{r}) \dd{\vect{r}} $ \\
 ${\color{white}\fn{\tilde{\chi}_{\vect{f}}}{\vect{k}} } = \rho \expval{\abs{\vect{f}}^2} + \rho^2 \tilde{h}_v(\vect{k}) ~~~~~~$
 }
 &
 $ \frac{\qty[\text{weight}]^2}{[\text{volume}]}$
 &
 $
 \begin{cases}
  0, & \text{if hyperuniform}\\
 >0, & \text{if nonhyperuniform}
 \end{cases}
 $
 &
 $\rho \expval{\abs{\vect{f}}^2}$
 &
 $\fn{\tilde{\chi}}{\vect{k}}$
 \\

 \hline 
 \makecell{
  Structure factor,
 \\
 $\fn{S_{\vect{f}}}{\vect{k}} \equiv \lim_{N\to\infty} N^{-1} \E{\fn{\tilde{n}^*_{\vect{f}}}{\vect{k}} \Bigcdot \fn{\tilde{n}_{\vect{f}}}{\vect{k}}} $
 \\
 ${\color{white}\fn{\cal S_{\vect{f}}}{\vect{k}} \equiv} = \fn{\tilde{\chi}_{\vect{f}}}{\vect{k}}/\rho + (2\pi)^d \rho \abs{\expval{\vect{f}}}^2 \fn{\tilde{\delta}}{\vect{k}}~~~~~$}
 &
 $\qty[ \text{weight}]^2 $
 &
 $(2\pi)^d \rho \abs{\expval{\vect{f}}}^2 \fn{\tilde{\delta}}{\vect{k}}+\fn{\tilde{\chi}_{\vect{f}}}{\vect{0}}/\rho$
 &
 $\expval{\abs{\vect{f}}^2}$
 &
 $S(\vect{k})$
 \\  
 \hline
 \end{tabular}

\end{table*}
\endgroup
\bigskip

We define the {\it generalized total correlation} function associated with vector weights ${\bf f}_1$
and ${\bf f}_2$ separated by the displacement vector $\bf r$ to be 
\begin{equation}
h({\bf r};{\bf f}_1,{\bf f}_2) \equiv \frac{\rho_2({\bf r};{\bf f}_1,{\bf f}_2)}{\rho^2} - p({\bf f}_1) p({\bf f}_2).
\label{h}
\end{equation}
Note that, when there is no long-range order, $h({\bf r};{\bf f}_1,{\bf f}_2)$ tends to zero in the limit $|{\bf r}|\to \infty$.
Substitution of Eq. (\ref{h}) into Eq. (\ref{chi-r-1}) yields the final form of the weight-averaged autocovariance 
function in the thermodynamic limit:
\begin{equation}
 \CHI{\vect{f}}{\vect{r}} = \rho \E{\abs{\vect{f}}^2} \fn{\delta}{\vect{r}}
 + 
 \rho^2 \fn{h_{\vect f}}{\vect{r}},
\label{chi-r-2}
\end{equation}
where $h_\vect{f}({\bf r})$ is the {\it weight-averaged} total correlation function defined by
\begin{equation}
h_\vect{f}({\bf r}) \equiv g_{2,{\bf f}} - |\langle {\bf f} \rangle|^2,
\label{h_f}
\end{equation}
where $g_{2,{\bf f}}$ is the pair correlation functions defined by Eq. (\ref{eq:g_2f}).
Table \ref{tab:direct} summarizes the definitions of the weight-averaged correlation functions in direct space and their behaviors at the extreme values of their arguments.

The corresponding {\it weight-averaged spectral density} is obtained by taking the Fourier transform of relation (\ref{chi-r-2}):
\begin{equation}
 \fn{\tilde{\chi}_\vect{f}}{\vect{k}}
 = 
 \rho \E{\abs{\vect{f}}^2} + 
 \rho^2 \fn{\tilde{h}_\vect{f}}{\vect{k}}.
\label{spec-1}
\end{equation}
To quantify the change in spectral density due to particle weights $\vect{f}$ relative to the unweighted configurations, we
introduce the \textit{excess spectral density}, which is defined as 
\begin{align} \label{eq:ex-chif}
 \tilde{\chi}_\vect{f}^{\mathrm{(ex)}}({\bf k})
 \equiv
 \tilde{\chi}_\vect{f}({\bf k}) - \expval{\abs{\vect{f}}^2}\tilde{\chi}({\bf k}),
\end{align}
where $\tilde{\chi}({\bf k})$ is the unweighted spectral density.
Thus, since $\tilde{\chi}_\vect{f}({\bf k})$ tends to $\rho\expval{\abs{\vect{f}}^2}$ in the large-$|{\bf k}|$ limit, 
the excess spectral density tends to zero, i.e., $\lim_{|{\bf k}|\to\infty} \tilde{\chi}_\vect{f}^{\mathrm{(ex)}}({\bf k})=0$.
As shown in Appendix \ref{general-S}, the ensemble average of the weight-averaged structure factor $\fn{\cal S_{\vect{f}}}{\vect{k}}$ in the thermodynamic limit  is directly related to the spectral density $\tilde{\chi}_\vect{f}({\bf k})$ via the following relation:
\begin{equation} \label{eq:Sofk_weighted}
\fn{S_{\vect{f}}}{\vect{k}}=\fn{\tilde{\chi}_{\vect{f}}}{\vect{k}}/\rho, 
\end{equation}
where the forward-scattering contribution to $\fn{\cal S_{\vect{f}}}{\vect{k}}$ is removed.
Table \ref{tab:Fourier} summarizes the definitions of the weight-averaged spectral functions in Fourier space and their behaviors
at the extreme values of their arguments.

\noindent{\bf Remarks:}
\begin{enumerate}

\item Hyperuniformity for weighted point configurations is precisely defined in Sec. \ref{hyper} via the generalized spectral density (\ref{spec-1})
as well as the generalized local variance described in Sec. \ref{variance}. There, we will utilize the excess spectral density, defined
by Eq. (\ref{eq:ex-chif}), to quantify the change in the exponents associated with small-$|{\bf k}|$ spectral densities
in going from unweighted to weighted point configurations.

\item We see that for a special case in which each weight is a unit scalar, we recover from (\ref{chi-r-2}) and (\ref{spec-1}) the corresponding standard autocovariance function and spectral density
for unweighted points \cite{To03a},
which are given, respectively, by
\begin{equation}
\chi({\bf r}) = \rho \delta({\bf r}) +\rho^2 h({\bf r}),
\end{equation}
\begin{equation}
{\tilde \chi}({\bf k}) = \rho S({\bf k}) \equiv \rho [1 +\rho {\tilde h}({\bf k})],
\end{equation}
where $h({\bf r})$ and $S({\bf k})$ are 
the total correlation function and structure factor (without the forward-scattering contribution), respectively.

\item We can immediately obtain the appropriate equations for scalar weights $f_j$
by direct substitution of $\vect{f}_j=f_j$ into (\ref{chi-r-2}) and (\ref{spec-1}), respectively yielding the 
expressions for the autocovariance function
\begin{equation}
 \CHI{f}{\vect{r}} = 
\rho \E{f^2} \delta({\bf r}) + \rho^2 h_{f}({\bf r})
\label{chi-r-4}
\end{equation}
and the spectral density 
\begin{equation}
{\tilde \chi}_f({\bf k})= \rho \E{f^2} + \rho^2 {\tilde h}_f({\bf k}).
\label{spec-3}
\end{equation}
Note that Eqs. \eqref{chi-r-4} and \eqref{spec-3} can also be obtained when 
the weights are vectors of different magnitudes but aligned in the same direction.

\item 
When the weights are uncorrelated and their means are identically zero, 
the generalized total correlation \eqref{h} can be reduced to a product of the total correlation of point centers and probability density functions of weights:
\begin{align*}
 \fn{h}{\vect{r};\vect{f}_1,\vect{f}_2} = \fn{h}{\vect{r}} \fn{p}{\vect{f}_1}\fn{p}{\vect{f}_2},
\end{align*}
and then the weight-averaged total correlation function \eqref{h_f} becomes zero for any $\vect{r}$, i.e., 
\begin{eqnarray}\label{eq:h=0}
 \weighted{h}{\vect{f}}{\vect{r}} &=& \fn{h}{\vect{r}} \qty [\int \vect{f}_1^* \fn{p}{\vect{f}_1}\dd{\vect{f}_1}] \Bigcdot \qty [\int \vect{f}_2 \fn{p}{\vect{f}_2}\dd{\vect{f}_2}] \nonumber\\
&=& 0,
\end{eqnarray}
implying that the weight-averaged autocovariance function, defined by Eq. (\ref{chi-r-2}), is given by
\begin{equation}
 \CHI{\vect{f}}{\vect{r}} = \rho \E{\abs{\vect{f}}^2} \fn{\delta}{\vect{r}}.
\label{chi-r-3}
\end{equation}
This outcome is the analog of what occurs in the case of unweighted Poisson point processes, for which
the standard total correlation is identically zero for all $\bf r$.
Analogously, the corresponding weight-averaged spectral density takes a constant value, i.e., $\tilde{\chi}_\vect{f}(\vect{k})=\rho \expval{\abs{\vect{f}}^2}$ for all $\vect{k}$, as shown later in Fig. \ref{fig:2D-RSA-shuffled}(b).

\item 
When the operator $\Bigcdot$ represents a dyadic product, the autocovariance function $\CHI{\vect{f}}{\vect{r}}$ given in Eq. \eqref{autoco} becomes a second-rank tensor.
Note that the trace of this tensor is the same as the autocovariance function when $\Bigcdot$ represents an inner product.
\item In the special case when the weight $\bf f$ takes {\it discrete} values, formula (\ref{chi-r-1}) for the autocovariance function
reduces to the following expression:
\begin{align}
\CHI{\vect{f}}{\vect{r}} =& \rho \E{\abs{\vect{f}}^2} \fn{\delta}{\vect{r}}
 + 
 \sum_{{\bf f}_1}  \sum_{{\bf f}_2} ({\bf f}_1^*\Bigcdot{\vect{f}}_2) 
 \nonumber \\
  &\times 
 \, \qty [\fn{\rho_2}{\vect{r};\vect{f}_1,\vect{f}_2} - \fn{\rho_1}{\vect{r}_1;\vect{f}_1} \fn{\rho_1}{\vect{r}_2;\vect{f}_2}].
\label{chi-r-3-special}
\end{align}
Here $\rho_1({\bf r};{\bf f})$ and $\fn{\rho_2}{\vect{r};\vect{f}_1,\vect{f}_2}$ are now probability distributions (not densities) with respect to the weights.
While the more compact form (\ref{chi-r-2}) for the autocovariance still applies, the weight-averaged total correlation function
is given by
\begin{equation}
h_\vect{f}({\bf r}) \equiv  \sum_{{\bf f}_1}  \sum_{{\bf f}_2} ({\bf f}_1^*\Bigcdot{\bf f}_2)
 \, h({\bf r};{\bf f}_1,{\bf f}_2).
\label{h_f-special}
\end{equation}

\end{enumerate}

\section{Local Variance}
\label{variance}

We now derive formulas for the {\it weight-averaged  local variance} associated with vector-weighted point configurations for a spherical window of radius $R$ in $d$-dimensional Euclidean space $\mathbb{R}^d$. The vector nature of the weights
allows for different definitions of the local variance, whether abstractly defined or dictated by the physics
of the problem. Following Torquato and Stillinger \cite{To03a}, we introduce
the following window indicator function for a spherical window of radius $R$ centered at position ${\bf x}_0$:
\begin{equation}
w({\bf x}-{\bf x}_0;{R})
=
\begin{cases}
  1, & \abs{\vect{x}-\vect{x}_0} \leq R, \\
  0, & \abs{\vect{x}-\vect{x}_0} > R.
\end{cases}
\label{window}
\end{equation}
To begin, we choose to define the sum of the vector weights of points $\weighted{{\cal N}}{\vect{f}}{R}$ within a window of radius $R$ by
\begin{align}
 \weighted{{\cal N}}{\vect{f}}{R} 
 \equiv& 
 \int \field{\vect{f}}{\vect{x}}
w({\bf x}-{\bf x}_0;{R}) \dd {\bf x} 
\nonumber \\
=& 
\sum_{j=1}^N \int {\bf f}_ j \fn{\delta}{\vect{x}-\vect{r}_j} \fn{w}{\vect{x}-\vect{x}_0; R} \dd{\bf x} 
\nonumber \\
=& 
\sum_{j=1}^N {\bf f}_j \fn{w}{{\bf r}_j-{\bf x}_0;{R}}.
\label{N}
\end{align}

The mean of $\weighted{{\cal N}}{\vect{f}}{R}$ is given by
\begin{align}
 \E{\weighted{{\cal N}}{\vect{f}}{R}}
 =& 
 \E{ \sum_{j=1}^N {\bf f}_j \fn{w}{\vect{r}_j - \vect{x}_0;R} }
 \nonumber \\
 =& 
 \int \fn{\rho_1}{\vect{Q}_1} {\bf f}_1 \fn{w}{{\bf r}_1-{\bf x}_0;R} \dd{\vect{f}_1}\dd{\bf r}_1 
 \nonumber \\
 =& 
 \rho \int {\bf f}_1 \fn{p}{{\bf f}_1} \ \fn{w}{{\bf r}_1-{\bf x}_0;R} \dd{\bf f}_1\,\dd{\bf r}_1 
 \nonumber \\
 =& 
 \rho \E{\bf f} \int w({\bf r};R) \dd{\bf r} 
 = \rho \E{ \bf f} v_1(R),
\label{N(R)}
\end{align}
where 
\begin{equation}
v_1(R)= \frac{\pi^{d/2} R^d}{\Gamma(1+d/2)}
\label{v1}
\end{equation}
is the volume of a $d$-dimensional spherical window of radius $R$. 
Note the possibility that the average $\E{\weighted{{\cal N}}{\vect{f}}{R}}$ is zero.

\begin{widetext}
We can generalize the second moment of $\weighted{{\cal N}}{\vect{f}}{R}$ to be the average of the dyadic (second-rank tensor) involving the product of the vector $\weighted{{\cal N}}{\vect{f}}{R}$, defined by Eq. \eqref{N}, with itself.
For simplicity, we consider to define the second moment to be the average of the inner product of the vector $\weighted{{\cal N}}{\vect{f}}{R}$ with itself: 
\begin{align}
  &\E{\weighted{{\cal N}^*}{\vect{f}}{R}\Bigcdot \weighted{{\cal N}}{\vect{f}}{R}}
 =
 \E {\sum_{j=1}^N |{\bf f}_j|^2 w({\bf r}_j-{\bf x}_0;R) + \sum_{j\neq k}^N  ({\bf f}_1^*\Bigcdot{\bf f}_2) \,
 \fn{w}{{\bf r}_j-{\bf x}_0;R} \fn{w}{{\bf r}_k-{\bf x}_0;R} } 
 \nonumber \\
 =& 
 \int |{\bf f}_1|^2 \fn{\rho_1}{{\bf Q}_1}  \fn{w}{{\bf r}_1-{\bf x}_0;R} \dd{\bf Q}_1 
 +\int ({\bf f}_1^*\Bigcdot{\bf f}_2) \,\rho_2({\bf Q}_1,{\bf Q}_2) \fn{w}{{\bf r}_1-{\bf x}_0;R} \fn{w}{{\bf r}_2-{\bf x}_0;R} \dd{\bf Q}_1 \dd{\bf Q}_2
 \nonumber \\
 =& 
 \rho \int |{\bf f}_1|^2 p({\bf f}_1) \fn{w}{{\bf r}_1-{\bf x}_0;R} \,\dd{\bf f}_1 
 +
 \int ({\bf f}_1^*\Bigcdot{\bf f}_2) \,\rho_2({\bf r}_1,{\bf r}_2;{\bf f}_1,{\bf f}_2 )w({\bf r}_1-{\bf x}_0;R) w({\bf r}_2-{\bf x}_0;R)\, \dd{\bf f}_1 \dd{\bf f}_2 \dd{\bf r}_1 \dd{\bf r}_2
 \nonumber\\
 =& 
 \rho \E{|{\bf f}|^2} v_1(R) 
 +
 \int ({\bf f}_1^*\Bigcdot{\bf f}_2) \,\rho_2({\bf r};{\bf f}_1,{\bf f}_2 )\fn{w}{{\bf r}_1-{\bf x}_0;R} \fn{w}{{\bf r}_2-{\bf x}_0;R}\, \dd{\bf f}_1 \dd{\bf f}_2 \dd{\bf x}_0 \dd{\bf r},
\label{second}
\end{align}
where $\weighted{{\cal N}^*}{\vect{f}}{R}\equiv [\weighted{{\cal N}}{\vect{f}}{R}]^*$.

Therefore, using relations (\ref{N(R)}) and (\ref{second}) yields the local variance to be
\begin{align}
 \weighted{\sigma^2}{\vect{f}}{R} 
 \equiv& 
 \E{ \weighted{{\cal N}^*}{\vect{f}}{R} \Bigcdot \weighted{{\cal N}}{\vect{f}}{R}} - \E{\weighted{{\cal N}^*}{\vect{f}}{R}} \Bigcdot \E{\weighted{{\cal N}}{\vect{f}}{R}} 
 \nonumber\\
 =& 
 \rho \E{|{\bf f}|^2} v_1(R) + \int ({\bf f}_1^*\Bigcdot{\bf f}_2) \,[ \rho_2({\bf r};{\bf f}_1,{\bf f}_2 )-
 \rho^2 p({\bf f}_1) p({\bf f}_2) ] w({\bf r}_1-{\bf x}_0;R) w({\bf r}_2-{\bf x}_0;R)\, \dd{\bf f}_1 \dd{\bf f}_2 \dd{\bf r}
 \nonumber \\
 =& 
 \rho \E {|{\bf f}|^2} v_1(R) 
 + \rho^2 v_1(R)\int_{\mathbb{R}^d} h_\vect{f}({\bf r}) \alpha_2(r;R)\, \dd{\bf r} 
 \nonumber\\
 =&
 v_1(R) \int_{\mathbb{R}^d} \CHI{\vect{f}}{\vect{r}} 
 {\alpha}_2(r;R)  \dd{\bf r} ,
\label{var-direct}
\end{align}
where $\CHI{\vect{f}}{\vect{r}}$ and $h_\vect{f}({\bf r})$ are given by Eqs. \eqref{chi-r-2} and \eqref{h_f}, respectively, and $\alpha_2(r;R)$ is the intersection volume of two identical
hyperspheres of radius $R$ (scaled by the volume of a sphere) whose centers
are separated by a distance $r$. 
\end{widetext}
The latter is known explicitly in any space dimension in various representations \cite{To03a,To06b}.
In the first three space dimensions, we have 
\begin{align}  \label{eq:alpha_direct}
  &\alpha_2(r; R) \nonumber \\
 = 
  &\Theta(1-x)
 \begin{cases}
    1 - x, \quad &d = 1, \\
 \frac{2}{\pi} \left[ \cos^{-1} \left( x \right) - x \sqrt{1 - x^2} \right], & d = 2, \\
    1 - \frac{3}{2}x + \frac{1}{2} x^3, &d = 3,
 \end{cases}
\end{align}
where $x\equiv r/2R$, and $\Theta(x)$ (equal to 1 for $x>0$ and zero otherwise) is the Heaviside step function.

Applying Parseval's theorem in the last line of Eq. (\ref{var-direct}) and using Eq. (\ref{spec-1}) yields the corresponding Fourier
representation of the local variance: 
\begin{equation}
 \weighted{\sigma^2}{\vect{f}}{R} = 
\frac{v_1(R) }{(2\pi)^d} \int_{\mathbb{R}^d} \weighted{\tilde{\chi}}{\vect{f}}{\vect{k}} 
 {\tilde \alpha}_2(k;R)  \dd{\bf k} .
\label{var-Fourier}
\end{equation}
Here $\weighted{\tilde{\chi}}{\vect{f}}{\vect{k}}$ is given by Eq. (\ref{spec-1}) and ${\tilde \alpha}_2(k;R)$ is the Fourier transform of $\alpha_2(r;R)$, which is known explicitly in
any space dimension in terms of Bessel functions $J_\nu(x)$ of order $\nu$ of the first kind \cite{To03a}; specifically,
\begin{equation} \label{eq:alpha-fourier}
 \tilde{\alpha}_2(k;R) = 2^d \pi^{d/2} \Gamma(1+d/2)\frac{[J_{d/2}(kR)]^2}{k^d}.
\end{equation}

\bigskip

\noindent{\bf Remarks:}
\begin{enumerate}
\item The nature of the operator $\Bigcdot$ determines the tensor rank of $\weighted{\sigma^2}{\vect{f}}{R}$.
If we take this operator to represent an inner product, $\weighted{\sigma^2}{\vect{f}}{R}$, defined by Eq. \eqref{var-direct}, becomes a scalar quantity identical to a sum of the variances $\weighted{\sigma^2}{[\vect{f}]_i}{R}$ in the individual components of $\vect{f}$ because 
\begin{align*}
 \weighted{\sigma^2}{\vect{f}}{R} 
 \equiv &
 \E{\weighted{{\cal N}^*}{\vect{f}}{R} \Bigcdot \weighted{{\cal N}}{\vect{f}}{R}} - \E{\weighted{{\cal N}^*}{\vect{f}}{R}}\Bigcdot\E{\weighted{{\cal N}}{\vect{f}}{R}} 
  \\
 =& 
 \sum_{i=1}^d \qty[\E{ \weighted{{\cal N}^*}{[\vect{f}]_i}{R} \weighted{{\cal N}}{[\vect{f}]_i}{R}} - \E{\weighted{{\cal N}^*}{[\vect{f}]_i}{R}} \E{\weighted{{\cal N}}{[\vect{f}]_i}{R}}]\\
 =&
 \sum_{i=1}^d \weighted{\sigma^2}{[\vect{f}]_i}{R},
\end{align*}
where $[\vect{x}]_i$ indicates the $i$th component of a vector $\vect{x}$.
By contrast, if $\Bigcdot$ represents a dyadic product, $\weighted{\sigma^2}{\vect{f}}{R}$ becomes a second-rank tensor identical to the covariance matrix of a random vector $\weighted{{\cal N}}{\vect{f}}{R}$ because
\begin{align*}
 [\weighted{\sigma^2}{\vect{f}}{R}]_{ij} 
 \equiv &
 \qty[\E{\weighted{{\cal N}^*}{\vect{f}}{R} \Bigcdot \weighted{{\cal N}}{\vect{f}}{R}} - \E{\weighted{{\cal N}^*}{\vect{f}}{R}}\Bigcdot\E{\weighted{{\cal N}}{\vect{f}}{R}}]_{ij} 
  \\
 =& 
 \qty[\E{ \weighted{{\cal N}^*}{[\vect{f}]_i}{R} \weighted{{\cal N}}{[\vect{f}]_j}{R}} - \E{\weighted{{\cal N}^*}{[\vect{f}]_i}{R}} \E{\weighted{{\cal N}}{[\vect{f}]_j}{R}}], 
\end{align*}
for $i,j=1,\ldots,d$.

\item In Sec. \ref{hyper}, we will use the Fourier
representation of the local variance (\ref{var-Fourier}) to obtain the conditions under which its large-$R$ asymptotic behavior
defines different hyperuniformity classes as well as nonhyperuniformity.

\item For the trivial case in which the weights are all unit scalars, we recover from (\ref{var-direct}) and (\ref{var-Fourier}) the corresponding expressions for the local number variance
of standard unweighted point processes given in Ref. \cite{To03a}.

\item For the Poisson counterpart of weighted point processes (i.e., uncorrelated weights with zero means), the local variance \eqref{var-direct} 
grows like the window volume, i.e., $\weighted{\sigma^2}{\vect{f}}{R} = \rho \E{\abs{\vect{f}}^2} \fn{v_1}{R}$,
which is the typical nonhyperuniform behavior, as detailed in Sec. \ref{hyper}.

\item We can immediately obtain the appropriate direct- and Fourier-space representations of the local variances for scalar weights $f_j$
by direct substitution of $\vect{f}_j=f_j$ into (\ref{var-direct}) and (\ref{var-Fourier}), yielding
\begin{align}
\sigma_f^2(R)
=&
\rho v_1(R)\Big[ \langle f^2\rangle + \rho \int \weighted{h}{f}{\vect{r}} \alpha_2(r;R)\, \dd{\bf r}\Big],
\label{var-direct-3}\\
=&\frac{ v_1(R)}{(2\pi)^d} \int_{\mathbb{R}^d} {\tilde \chi}_f({\bf k}) 
{\tilde \alpha}_2(k;R)  \dd{\bf k}.
\label{var-Fourier-3}
\end{align}

\end{enumerate}

\section{Hyperuniformity and Nonhyperuniformity}
\label{hyper}

Following the same analysis as in Ref. \cite{To18a}, it is simple to show
that dividing the local variance (\ref{var-Fourier}) by the window volume $ v_1(R)$ 
and taking the limit $R \to \infty$ yields
\begin{equation}
\lim_{R \rightarrow \infty}\frac{\displaystyle \sigma^2_{\bf f}(R)}{\displaystyle v_1( R)}= \lim_{|{\bf k}| \to 0}{\tilde \chi}_{\vect f}({\bf k})
 = \int_{\mathbb{R}^d}\CHI{\vect{f}}{\vect{r}} \dd {\vect{r}}.
\label{VAR1}
\end{equation}
We can define hyperuniformity for weighted point configurations via the generalized spectral density (\ref{spec-1})
in the usual way for general random variables \cite{To16a}, namely, the hyperuniformity condition
is given by
\begin{equation}
\lim_{|{\bf k}| \to 0}{\tilde \chi}_{\vect f}({\bf k})=0,
\label{condition-1}
\end{equation}
and, hence,
\begin{equation}
\lim_{R\rightarrow \infty}\frac{\displaystyle \sigma^2_{\bf f}(R)}{\displaystyle v_1(R)}= 0,
\label{VAR2}
\end{equation}
implying that the variance for a hyperuniform system grows more slowly than the
window volume. Equations (\ref{VAR1}) and (\ref{VAR2}) yield the direct-space hyperuniformity sum rule 
\begin{equation}
 \int_{\mathbb{R}^d}\CHI{\vect{f}}{\vect{r}} \dd{\vect{r}}= 0
\label{sum-rule}
\end{equation}
or, equivalently,
\begin{equation}
\rho \int_{\mathbb{R}^d} \weighted{h}{\vect{f}}{\vect{r}} \dd {\vect{r}} =- \langle |{\bf f}|^2 \rangle.
\label{condition-2}
\end{equation}
We infer from this sum rule that in order to get hyperuniformity, negative correlations are required.
Note that when the weights are uncorrelated, and their means
are identically zero, the system cannot be hyperuniform, since
the sum rule above cannot be satisfied, i.e.,
\begin{equation}
\int_{\mathbb{R}^d}\CHI{\vect{f}}{\vect{r}} = \rho \E{\abs{\vect{f}}^2},
\label{uncorr}
\end{equation}
where we have used Eq. (\ref{chi-r-3}).

We can obtain the large-$R$ asymptotic expansion of the weighted local variance by substituting the first two terms  
of the small-$r$ series expansion of the scaled intersection volume $\alpha_2(r; R)$ into Eq. (\ref{var-direct}), as was done for the unweighted point configurations procedure laid out in Ref. \cite{To03a}, yielding 
\begin{align}
\label{eq:sigma_series}
 \sigma_{\bf f}^2(R) = 2^d \phi & \left[ A_{\bf f}(R) \left( \frac{R}{D} \right)^d + B_{\bf f}(R) \left( \frac{R}{D} \right)^{d-1} \right. \nonumber\\
  &\left. + o\left( \left( \frac{R}{D} \right)^{d-1} \right) \right] \quad (R \to \infty),
\end{align}
where $D$ is a characteristic microscopic length scale ({\it e.g.}, mean nearest-neighbor distance), and $\phi = \rho v_1(D/2)$ is a dimensionless density.
The coefficient $A_{\bf f}(R)$ is a $d$-dependent asymptotic coefficient that multiplies terms proportional to the window volume and is obtained from the autocovariance function as
\begin{equation}
 A_{\bf f}(R) = \frac{1}{\rho} \int_{|\mathbf{r}| \leq 2R} \CHI{\vect{f}}{\vect{r}} \, d\mathbf{r}.
\end{equation}
The coefficient $B_{\bf f}(R)$ represents the window surface area and is given by
\begin{equation}
 B_{\bf f}(R) = -\frac{\, c(d)}{2D \, \rho} \int_{|\mathbf{r}| \leq 2R} \CHI{\vect{f}}{\vect{r}} |\mathbf{r}| \, d\mathbf{r},
\end{equation}
with the coefficient $c(d)$ given by
\begin{equation}
 c(d) = \frac{2\Gamma(1+d/2)}{\pi^{1/2} \Gamma[(1+d)/2]}.
\end{equation}

In the $R \to \infty$ limit, it is clear that the volume coefficient $A_{\bf f}(R)$ is equal to the spectral density $\weighted{\tilde{\chi}}{\vect{f}}{\vect{k}}$ at zero wave number, {\it i.e.},
\begin{equation}
\label{eq:A_N}
 \overline{A}_{\bf f} \equiv \lim_{R \to \infty} A_{\bf f}(R) = \lim_{|\mathbf{k}| \to 0} \weighted{\tilde{\chi}}{\vect{f}}{\vect{k}}  \geq 0.
\end{equation}
Following the similar analysis as in Ref. \cite{To21c}, the surface coefficient in the same limit can be expressed as a convergent constant
\begin{align} 
 \overline{B}_\vect{f} \equiv \lim_{R\to\infty} B_\vect{f} (R) ,
  \label{eq:B_f}
\end{align}
for disordered systems in which the autocovariance function decays sufficiently fast to yield a convergent integral.
These include typical nonhyperuniform systems
as well as the strongest form of hyperuniform defined as class I below.
For class I systems, we can define the hyperuniformity order metric $\overline{\Lambda}_\vect{f}$, proportional to $\overline{B}_\vect{f}$ to rank order such hyperuniform systems:
\begin{align}
 \overline{\Lambda}_\vect{f} 
 \equiv & \lim_{ L \to \infty } \frac{1}{L} \int_{0}^L \frac{\sigma_{\vect{f}}^2 (R) }{(R/D)^{d-1}} \dd{R} 
 = 2^d \phi \overline{B}_\vect{f}.
 \label{eq:B} 
 \end{align}
Following Ref. \cite{To21c}, it is easy to obtain
the equivalent Fourier representation
of this order metric:
\begin{align}
 \overline{\Lambda}_\vect{f}  =& \frac{v_1(D)}{D} \frac{d}{\pi}\int_0 ^\infty \frac{\tilde{\chi}_\vect{f}(k)-\tilde{\chi}_\vect{f}(0)}{k^2} \dd{k}. \label{eq:Lambda-exact}
\end{align}
A practically useful numerical implementation of Eq. \eqref{eq:Lambda-exact} is provided in Appendix \ref{app:Lambda_f}.

We see from relation (\ref{var-Fourier}) that if the spectral density $\weighted{\tilde{\chi}}{\vect{f}}{\vect{k}}$ tends to zero in the limit
$|{\bf k}| \to 0$, i.e., obeys the hyperuniformity condition (\ref{condition-1}), the weighted local variance $\weighted{\sigma^2}{\vect{f}}{R}$ grows more slowly than $R^d$, 
and, hence, the growth laws do not change from those in unweighted point configurations, including the established three hyperuniformity classes: I, II, and III \cite{To18a}.
Specifically, when the generalized spectral density goes to zero with the
following power-law scaling \cite{Za09,To16b,To18a}:
\begin{equation}
{\tilde \chi}_{\bf f}({\bf k})\sim |{\bf k}|^{\alpha_{\bf f}} \quad (|{\bf k}| \to 0),
\label{scaling}
\end{equation}
namely,
\begin{align}  
\sigma^2_{\bf f}(R) \sim 
\begin{cases}
R^{d-1}, \quad\quad\quad \alpha_{\bf f} >1 \qquad &\text{(Class I)}\\
R^{d-1} \ln R, \quad \alpha_{\bf f} = 1 \qquad &\text{(Class II)},\\
R^{d-\alpha_{\bf f}}, \quad 0 < \alpha_{\bf f} < 1\qquad &\text{(Class III)}
\end{cases}
\label{eq:classes}
\end{align}
where the exponent $\alpha_{\bf f}$ is a positive constant.
Classes I and III are the strongest and weakest forms of hyperuniformity, respectively.
We can extend the notion of stealthy hyperuniform unweighted point configurations to 
those with weights in the obvious way, namely, the latter
are defined to be those that possess
zero-scattering intensity for a set of wavevectors around the origin \cite{To16b}, i.e.,
\begin{equation}
{\tilde \chi}_{\bf f}({\bf k})=0 \qquad \mbox{for}\; 0 \le |{\bf k}| \le K.
\label{stealth}
\end{equation}
Such stealthy hyperuniform systems are of class I.

By contrast, for any nonhyperuniform weighted point configuration, it is straightforward to show,
using a similar analysis as for unweighted point configurations \cite{To21c}, that
the local variance has the following large-$R$ scaling behaviors:
\begin{align} 
\sigma^2_{\bf f}(R) \sim 
\begin{cases}
R^{d}, & \alpha_{\bf f} =0 \quad \text{(typical nonhyperuniform)}\\
R^{d-\alpha_{\bf f}}, & -d <\alpha_{\bf f} < 0 \quad \text{(antihyperuniform)}.\\
\end{cases}
\label{sigma-nonhyper}
\end{align}
For a ``typical" nonhyperuniform system, ${\tilde \chi}_{\bf f}(0)$ is bounded \cite{To18a}. By definition, {\it antihyperuniform} weighted point configurations
are those in which 
${\tilde \chi}_{\bf f}(0)$ is unbounded, i.e.,
\begin{equation}
\lim_{|{\bf k}| \to 0} {\tilde \chi}_{\bf f}({\bf k})=+\infty,
\label{antihyper}
\end{equation}
and, hence, are diametrically opposite to hyperuniform systems.
It is useful to note here that large-$|\bf r|$ asymptotic behavior of the autocovariance function $\chi_{\bf f}({\bf r})$
of an antihyperuniform system has the following inverse power-law form \cite{To18a}:
\begin{equation}
\chi_{\bf f}({\bf r}) \sim \frac{1}{|{\bf r}|^{d+\alpha_{\bf f}}} \qquad (|{\bf r}| \to \infty), 
\label{anti}
\end{equation}
where $\alpha_{\bf f}$ lies in the open interval (-$d$,$0$).

Interestingly, 
it is not necessarily true that if the unweighted point process is itself hyperuniform, then the weighted point process
is also hyperuniform. Moreover, if the underlying point configurations are nonhyperuniform, the corresponding weighted point processes
may {\it still} be hyperuniform. We will present specific examples of both scenarios in Secs. \ref{revisit} and \ref{volume}.
To quantify such changes in the large-scale structural behaviors, we utilize the excess spectral density $\tilde{\chi}_\vect{f}^\mathrm{(ex)}(\vect{k})$, defined in Eq. \eqref{eq:ex-chif}, to define the change in the hyperuniformity exponent $\delta \alpha$ as follows:
\begin{align}  \label{eq:Delta_alpha}
 \delta \alpha \equiv \alpha_\vect{f} - \alpha,
\end{align}
where $\alpha$ is the scaling exponent of the structure factor of the unweighted point configuration, i.e., $S({\bf k})\sim 
\tilde{\chi}({\bf k})\sim |{\bf k}|^\alpha$ in the limit $|{\bf k}|\to 0$.

\section{Revisiting Hexatic/Nematic/Tetratic Phases and Dipolar Systems Under the Hyperuniformity Lens}
\label{revisit}

We begin by revisiting previous well-established
results for bond-orientationally ordered (hexatic/nematic/tetratic) phases and dipolar systems under the hyperuniformity
lens of weighted point configurations. Subsequently, we carry out additional calculations to quantify
the weighted total correlation function $h_{\psi_6}({\bf r})$ associated with sixfold bond-orientational order 
in the solid phase of 2D hard disks as well as certain 2D hyperuniform packings.

\begin{figure*}[t]
\includegraphics[width=0.9\textwidth]{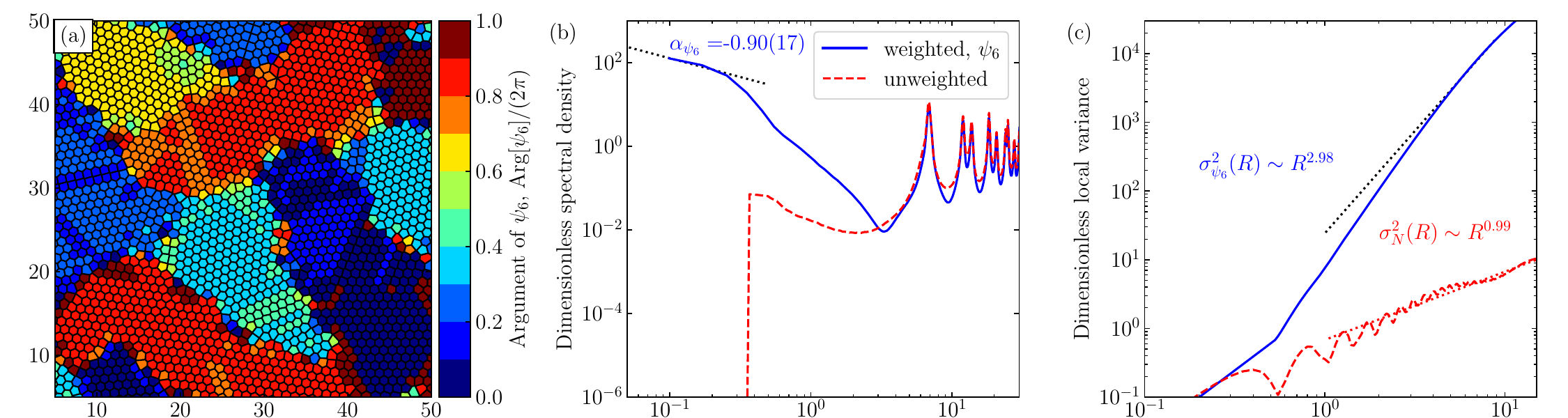}
\caption{
Point configurations derived from 2D ultradense hyperuniform packings with packing fraction $\phi=0.86$ \cite{To25a,Ki25a} weighted with the sixfold bond-orientational order parameter $\psi_6$.
(a) Representative image of a 2D weighted point configuration.
The colors depict the argument of the complex-valued parameter $\psi_6$. The high degree of polycrystallinity is clearly 
picked up by the 6-fold bond-orientational order.
(b) Log-log plot of the dimensionless spectral densities $\tilde{\chi}(k)/\rho$ and $\tilde{\chi}_{\psi_6}(k)/\rho$ for 2D unweighted and weighted point configurations.
The weighted one is antihyperuniform with $\alpha_{\psi_6}\approx -0.90\pm 0.17$.
(c) Log-log plot of the dimensionless local variances $\sigma_N^2(R)$ and $\sigma_{\psi_6}^2(R)$ for the unweighted and weighted ones, respectively.
The black and red dotted lines illustrate the large-$R$ growth rate of the local variances.
\label{fig:2D_SHU_psi6}
}
\end{figure*}

\subsection{2D bond-orientationally ordered, liquid, and solid phases}
\label{s:orientation}

In two dimensions, the well-known bond-orientational order parameter $\psi_n$, which reflects local
$n$-fold rotational order, is a special case of a weighted point configuration in which the weights
are complex numbers. Specifically, for a particle $j$, its $n$-fold bond-orientational order parameter is defined as \cite{Ko73,Ha78,Ne22, Ca21,Di18a, Marc96}
\begin{align} \label{eq:psi_n}
\psi_n(j) \equiv \frac{1}{N_{j}} \sum_{k=1}^{N_{j}} \exp(i n\theta_{jk}), \quad (n=2,4,6,\dots),
\end{align}
where $N_j$ is the number of near neighbors for particle $j$ (e.g., associated with its Voronoi cell), $\theta_{jk}$ is the angle of the bond
formed between particle $j$ and its $k$-th neighbor and a reference axis, and $n$ is an even positive integer.
By setting $n$ equal to 2, 4, 6 or 10, one seeks to detect nematic, tetratic, hexatic, and
pentagonal orientational order, respectively \cite{Ko73,Ha78,Ne22, Ca21,Di18a, Marc96}.
The corresponding weighted autocovariance function $\chi_{\psi_n}({\bf r})$, total correlation function $h_{\psi_n}({\bf r})$ and spectral density
$\tilde{\chi}_{\psi_n}({\bf k})$ are obtained from the definitions (\ref{chi-r-2}), (\ref{h_f}) and (\ref{spec-1}), respectively, where
${\bf f}=\psi_n$.

\begin{table*}[ht]
\caption{Nonhyperuniform/Hyperuniform classification of 2D isotropic liquid, anisotropic liquid, and solid equilibrium phases
without and with weights.
 Values of the associated hyperuniformity exponent $\alpha_{\psi_n}$ and changes in the exponent $\delta \alpha$ are listed. \label{tab:classes}}
\begin{tabular}{c|c |c | c}
\hline 
Phases & \makecell{Point configurations \\ without weights ($\alpha$)} & \makecell{Point configurations \\ with weights $\psi_n$ ($\alpha_{\psi_n}$)} & \makecell{Change in exponent, \\ $\delta \alpha \equiv \alpha_{\psi_n}-\alpha$}\\
\hline 
Liquid & \makecell{Typical nonhyperuniform \\ ($\alpha=0$)} & \makecell{Typical nonhyperuniform \\ ($\alpha_{\psi_n}=0$)} & 0\\
\hline 
\makecell{Anisotropic liquid \\ (nematic, tetratic, hexatic)} & 
\makecell{Typical nonhyperuniform \\ ($\alpha=0$)}& \makecell{Antihyperuniform \\ ($\alpha_{\psi_n}\in[-2,-1]$)\footnote{According to the KTHNY theory, $\alpha_{\psi_n}\in[-2,-1.75]$.}} & $[-2,-1]$\\
\hline 
Solid & \makecell{Typical nonhyperuniform \\ ($\alpha=0$)}& \makecell{Antihyperuniform \\($\alpha_{\psi_6}\approx -0.20$)
\footnote{We have estimated this value in the present work from a single very large hard-disk configuration with $\phi=0.720$ and $N = 1024^2 = 1048576$ taken from Ref. \cite{Ber11}}
 } & $-0.20$ \\
\hline 
\end{tabular}
\end{table*}

The KTHNY theory \cite{Ko73,Ha78,Ne79} predicts 
two continuous phase transitions as 2D solid melts to a liquid: a solid to a hexatic (anisotropic) liquid phase 
with orientational order but no long-range translation order 
and then a transition from the hexatic to the isotropic liquid phase.
In the notation of the present paper, it was shown that within the hexatic phase, the weighted total correlation function $h_{\psi_6}({\bf r})$ 
for large $|{\bf r}|$ decays as the following power law:
\begin{equation}
h_{\psi_n}({\bf r}) \sim \frac{1}{|{\bf r}|^{\eta_6}} \qquad |{\bf r}| \to \infty,
\label{hex}
\end{equation}
where the exponent $\eta_6$ lies in the interval ($0,1/4]$ \cite{Ha78}, indicating it possesses
{\it quasi-long-range sixfold bond-orientational order}, where $\eta_6=1/4$ corresponds
to the system approaching the transition to the isotropic liquid. 
Numerous computer simulations and experiments of a variety of 2D physical systems have reported hexatic phases, including 2D colloids \cite{Ke07,Sh09}, 2D hard disks \cite{Ber11}, and active matter \cite{Mar13,Di18a,Kl18a,Pa20},
among others.

 The form of the algebraic decay of $h_{\psi_6}({\bf r})$
in Eq. (\ref{hex}) implies the same large-$|{\bf r}|$ decay in the autocovariance function (\ref{anti}) with ${\bf f}=\psi_6$, and, hence,
the small-$k$ spectral density exponent $\alpha_{\psi_6}=\eta_6-2$, which is negative. Hence, we conclude
that the hexatic phase is antihyperuniform with respect to sixfold bond-orientational order.
Furthermore, note that the change in the exponent in going from the unweighted typical nonhyperuniform
states ($\alpha=0$) to weighted antihyperuniform configurations involves the exponent change $\delta \alpha= \eta_6-2$, where $\delta \alpha$ is defined in (\ref{eq:Delta_alpha}). Hence, within the hexatic phase, $\delta \alpha$ varies between -7/4 and -2.


Via Monte Carlo simulations, Frenkel and Eppenga \cite{Frenk85} demonstrated that a nematic phase in 2D equilibrium hard-needle systems existed between a solid phase and an isotropic liquid phase and showed the twofold correlation function exhibits quasi-long-range order of the form $g_{2,\psi_2}({\bf r}) \sim 1/|{\bf r}|^{\eta_2}$ for large $|\bf r|$;
see also Ref. \cite{Cu90} for similar findings in the case of hard ellipses.
Within the nematic phase, $\eta_2$ is expected to lie in the interval ($0,1/4$], implying that this phase
is antihyperuniform with respect to twofold bond-orientational order.
The exponent $\eta_2$ takes the value $1/4$ as the system approaches the transition to the isotropic liquid. 

Wojciechowski and Frenkel \cite{Wo04} demonstrated through Monte Carlo simulations that a tetratic phase can exist in 2D equilibrium hard-square systems.
They showed that the fourfold correlation function indicates quasi-long-range order of the form $ g_{2, \psi_4}({\bf r}) \sim |{\bf r}|^{-\eta_4}$ 
for large $|\bf r|$. Experiments \cite{Lo25} on the same systems support this finding. 
Similar findings were also reported by Donev et al. \cite{Do06a} in simulations of hard rectangles with an aspect ratio two,
where $\eta_4$ was estimated to be equal to 1 at the tetratic-liquid transition. 
All of these studies indicate that the tetratic phase is
antihyperuniform with respect to fourfold bond-orientational ordered.

The lack of positional order in bond-orientationally ordered phases (e.g., hexatic, tetratic, or nematic) implies that their unweighted point configurations
are typical nonhyperuniform states \cite{To18a}, i.e., the structure factor is positive and bounded
at zero wave number. It is well known that the corresponding unweighted point configurations of liquid phases in such phase diagrams
are also typical nonhyperuniform states \cite{To18a}, and hence the change in the exponent is $\delta\alpha=0$. 
Less well known is the fact that the corresponding unweighted point configurations 
within 2D solid phases away from jamming states are also typically nonhyperuniform; see, for example, the theoretical arguments
presented in Ref. \cite{To21c} for this conclusion in the case of hard-sphere solid phases in arbitrary dimension.

The bond-orientational order in the 2D solid phases is characterized by a positive global mean
value of the associated order parameter $\psi_n$, which remains
independent of the system size \cite{Ha78,Ne79}.
Because the precise manner in which $h_{\psi_n}({\bf r})$ decays asymptotically for large $r$ is not well
understood, we carry out here computations of $h_{\psi_6}({\bf r})$ to determine the corresponding 
the exponent $\alpha_{\psi_6}$ of a single but very large
configuration ($N = 1024^2 = 1048576$) of a 2D hard-disk solid with $\phi=0.720$, generated by Bernard and Krauth
\cite{Ber11}, yielded the exponent value $\alpha_{\psi_6}\approx -0.2$, implying that this state 
as well as the solid phase away from jamming is antihyperuniform with an exponent change of $\delta \alpha=-2$
(since $\alpha=0$).

In summary, for the solid, anisotropic liquid, and isotropic liquid phases of the 2D equilibrium systems, the changes in the hyperuniformity exponent in going from
unweighted to weighted configurations $\delta \alpha$, defined in Eq. (\ref{eq:Delta_alpha}), are $-0.2\;  (0.2-0)$, $-1 \; (-1-0)$, and $0 \; (0-0)$, respectively. Thus, the high degree
of bond order across length scales in both the anisotropic liquid and solid phases leads to antihyperuniform fluctuations with negative values
of $\delta \alpha$. This is to be contrasted with the isotropic
liquid phase in which bond-order fluctuations remain typically
nonhyperuniform such that $\delta \alpha=0$.
Table \ref{tab:classes} summarizes the nonhyperuniform and hyperuniform classification 
of 2D isotropic liquid, anisotropic liquid, and solid equilibrium phases
without and with weights given by $\psi_n$. We include in this table the associated hyperuniformity exponent $\alpha_{\psi_n}$ and changes in the exponent $\delta \alpha$.

Before closing this subsection, we present a novel case
in which we show that a 2D hyperuniform packing of circular disks
with respect to density fluctuations can be converted into a weighted {\it antihyperuniform} system with respect
to hexatic order. To our knowledge, this represents the first example of such an extreme change in the large-scale fluctuations in going from unweighted to weighted configurations.
 The unweighted hyperuniform packing 
is generated using the collective-coordinate optimization procedure to produce ``ultradense''
hyperuniform packings in the limit that the stealthiness parameter
$\chi$ tends to zero with a packing fraction $\phi=0.86$; see Refs. \cite{To25a, Ki25a}
for details. (Note that this procedure to generate
stealthy hyperuniform point configurations is different from the
one described in Sec. \ref{volume}.) Such packings are characterized by a high degree of polycrystallinity.
We find that the weight-averaged spectral density $\tilde{\chi}_{\psi_6}(k)$  shows a power-law divergence around the origin with a negative hyperuniformity exponent $\alpha_{\psi_6} \approx -0.90 \pm 0.17$; see Fig. \ref{fig:2D_SHU_psi6}(b).
Consistent with the theoretical formula (\ref{sigma-nonhyper}), such a trend in the spectral density implies that the associated local variance should grow like $R^{2.98}$, which is faster than the window volume growth ($R^2$); see Fig. \ref{fig:2D_SHU_psi6}(c).
Note that the local variance $\sigma_{\psi_6}^2(R)$ is also dimensionless, since the order parameter $\psi_6$ is dimensionless.

\begin{figure*}
\subfloat[]{\includegraphics[width=0.4\textwidth]{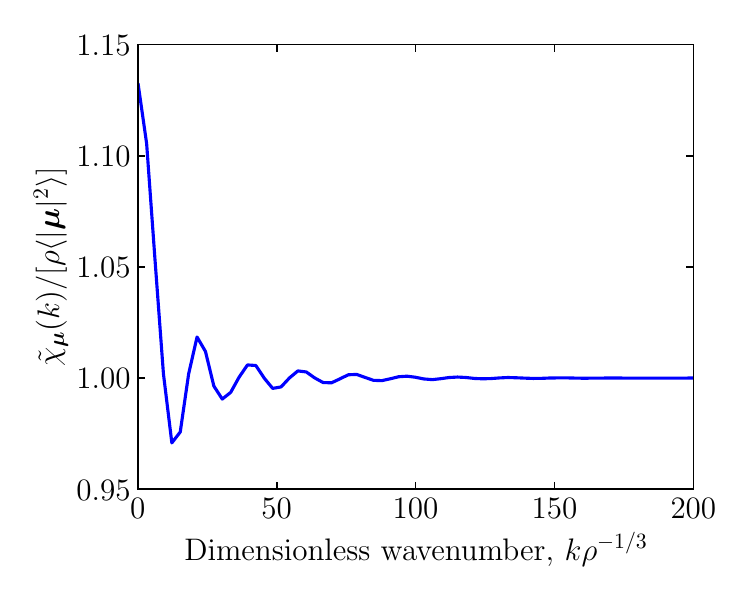}}
\subfloat[]{\includegraphics[width=0.4\textwidth]{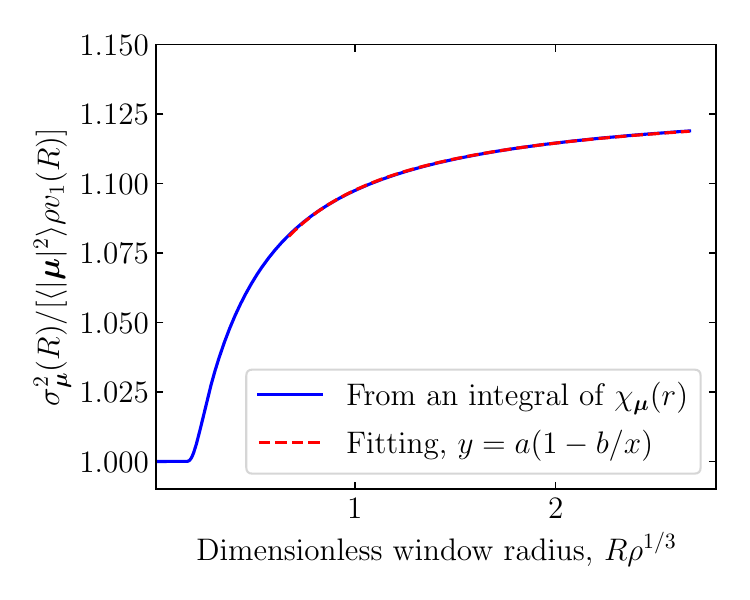}}

 \caption{
 Spectral density and local variance computed from the dipole correlation function $c_m(r)$ of an equilibrium water model in Ref. \cite{Sh07}.
 (a) Dimensionless generalized spectral density $\tilde{\chi}_{\bm \mu}(k) / [\rho \expval{\abs{\bm {\mu}}^2}]$ as a function of a wave number $k\rho^{-1/3}$.
 (b) Dimensionless local variance $\sigma_{\bm \mu} ^2(R)/ \expval{\abs{{\bm \mu}}^2}$ scaled by the dimensionless window volume $\rho v_1(R)$ as a function of a dimensionless window radius $R\rho^{1/3}$. 
  \label{fig:dipole}
 }
\end{figure*}

\subsection{Vector weights: reexamination of dipolar correlations in water}
\label{s:water}

A natural application of the formulas derived for the generalized weighted direct- and Fourier-space pair correlations  (Sec. \ref{auto-spec}) 
and local variance (Sec. \ref{variance}) is
to the study of systems in which the weights are
real-valued vectors that are prescribed by some probability
distribution $p({\bf f})$. Such situations include  dipolar systems \cite{Sh07}, spin vectors \cite{Ud21,Me84,Le19c}, bond vectors in polymer chains \cite{Fl69,Se10}, orientation vectors of anisotropic particles (e.g., liquid crystals \cite{St74,De95}) 
and velocity vectors in active particle systems \cite{Mar13}.

In this subsection, we 
 reexamine the simulation data of dipolar correlations in liquid water presented in Ref. \cite{Sh07} to assess whether they are hyperuniform or not. The large dielectric constant of water, which is related to the fluctuations of the dipole moments in such a macroscopic sample, is of great interest, since it is the 
most important polar solvent in chemistry and biology.
The dipole moments ${\bm \mu}$ of water molecules have similar average magnitudes, but their orientations are determined by the hydrogen-bond network \cite{Pa45}. To make contact with the correlation function used in Ref. \cite{Sh07}, we consider its relationship to
the weight-averaged autocovariance function $\chi_{{\bm \mu}}(\vect{r})$, defined by expression  (\ref{chi-r-2}), with ${\bf f}={\bm \mu}$.

This autocovariance function  is directly related to the dipole correlation function $c_m(\vect{r})$ employed in Ref. \cite{Sh07}:
\begin{align} \label{eq:chi_mu-c_m}
 \fn{\chi_{{\bm \mu}}}{\vect{r}} = \rho^2 c_m(\vect{r}) - \abs{\rho \expval{{\bm \mu}}}^2 + \rho \expval{\abs{\bm \mu}^2} \delta(\vect{r}) ,
\end{align}
where $c_m(\vect{r})$ is defined as 
\begin{align}
 c_m(\vect{r}) \equiv & \frac{1}{\rho} \qty[ \frac{1}{N} \int \dd{\vect{r}'} \expval{ n_{\bm \mu}(\vect{r}+\vect{r}') \cdot n_{\bm \mu}(\vect{r}')} - \expval{\abs{\bm \mu}^2} \delta(\vect{r}) ].
\end{align}
As per Ref. \cite{Sh07}, we take  the following values of parameters:  $\expval{\abs{\bm \mu}^2} \approx (3.09 ~D)^2 \approx 9.55 ~D^2$, $\expval{\bm \mu}\approx \vect{0}$ \footnote{It is inferred from the fact that $c_m(r) \to 0$ as $r\to \infty$.}, the number density $\rho \approx 4.456 \times 10^{3} D^2 \text{Bohr}^{-3}$, $D$ is the debye unit, and $\text{Bohr}=5.29...\times 10^{-11} m$ is the Bohr radius.

The estimated spectral density and local variance are shown in Fig. \ref{fig:dipole}. 
This figure clearly shows that dielectric dipole moments in liquid water in equilibrium are typically nonhyperuniform, as is its unweighted counterpart.
Specifically, as shown in Fig. \ref{fig:dipole}(a), the normalized spectral density tends to a bounded positive value as $k$ tends to zero.
This scaled function also converges to unity as the wave number increases greater than $k\rho^{-1/3}=100$, i.e., $\lim_{k\to\infty}\tilde{\chi}_{\bm{\mu}}(k)=\rho \expval{\abs{\bm {\mu}}^2}$, which is consistent with the prediction in Table \ref{tab:Fourier}.
Using Eq. \eqref{var-Fourier-3} and the spectral density shown in Fig. \ref{fig:dipole}(a), we also estimate the dimensionless local variance scaled by the window volume multiplied by  $\rho$, i.e.,  $\sigma_{\bm \mu} ^2(R)/ [\expval{\abs{{\bm \mu}}^2}\rho v_1(R)]$, as a function of
the dimensionless window radius $R\rho^{1/3}$; see Fig. \ref{fig:dipole}(b).
Since this system is typically nonhyperuniform, this scaled variance tends  to a positive constant as the window radius becomes large, implying that the local variance itself  tends to grow as fast as the window volume for sufficiently large window radii.

Note that if the autocovariance function, defined by Eq. (\ref{chi-r-2}), is a radial function, i.e., $\CHI{\bm{\mu}}{r}$, where $r=|\bf r|$, its volume integral over a spherical region radius $R$ is given by
\begin{align}  \label{eq:GK_R}
 G_K(R) \equiv \frac{d\pi^{d/2}}{\Gamma(1+d/2)} \int_{0}^R r^{d-1} \frac{\CHI{\vect{f}}{r}}{\langle |{\bf f}|^2\rangle} \dd{r}.
\end{align}
Thus, $G_K(R)$ can be viewed as a type of ``{\it cumulative coordination number}''
in the same way that a volume integral of the unweighted $\rho g_2({\bf r})$ of a point process
over such a spherical region is the cumulative coordination number $Z(R)$, equal to expected number
of points within a sphere of radius $R$.
For $d=3$, the function $G_K(R)$ is the well-known
{\it Kirkwood correlation function}.

\section{Voronoi-Cell Volume Weights}
\label{volume}

\begin{figure*}[t]
 \centering
 \includegraphics[width=0.95\textwidth]{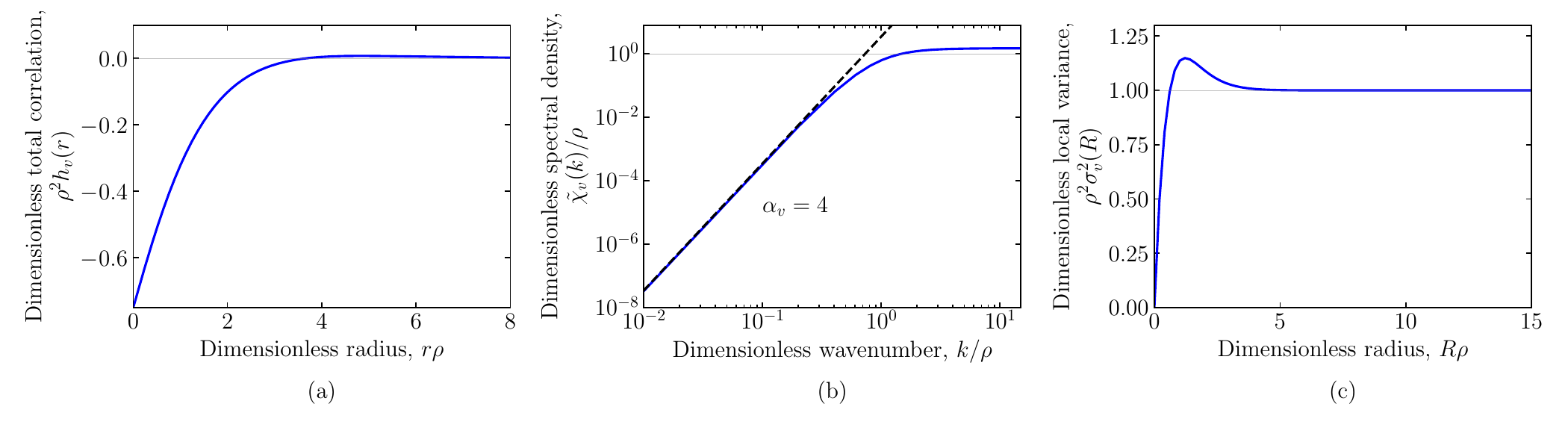}
 \caption{
 Theoretical prediction of a 1D Poisson point process weighted with the Voronoi volume.
(a) Dimensionless weight-averaged total correlation function $\rho^2 h_v(r)$ as a function of the dimensionless radius $r\rho$; see Eq. \eqref{eq:hv-1D-Poisson}.
(b) Dimensionless weight-averaged spectral density $\tilde{\chi}_v(k)$ as a function of the dimensionless wave number $k/\rho$ on a log-log scale; see Eq. \eqref{eq:chivk-1D-Poisson}.
(c) Dimensionless weight-averaged local variance $\rho^2 \sigma_v^2(R)$ as a function of the dimensionless radius $R\rho$ in a linear scale; see Eq. \eqref{eq:sigma-1D-Poisson}.
 }
 \label{fig:1D-Poisson-h}
\end{figure*}

Consider scalar weights $f_i$ that are the Voronoi-cell volumes $v_i$ of Voronoi tessellations of statistically homogeneous point configurations
in $\mathbb{R}^d$. We specifically obtain results for the weighted autocovariances, spectral densities and local variances
for certain 1D, 2D and 3D models of hyperuniform, standard nonhyperuniform and antihyperuniform unweighted point configurations 
and show that in each case the weighted point configurations are hyperuniform of class I. 
Such scalar-weighted point configurations can be viewed as first-order approximations of the two-phase hyperuniform sphere packings that were created from 2D and 3D random tessellations in Refs. \cite{Ki19a, Ki19b}.
These hyperuniform weighted point configurations also have been studied to model supercooled liquid \cite{Fa14} and 2D space-filling foams \cite{Ch21c}.
The class I hyperuniformity of these systems arises because, within a Voronoi cell, the weight (or volume) of the particle is identical to the cell volume. 
When a sufficiently large spherical window is placed, density fluctuations are measured only in the vicinity of the window boundary, where the Voronoi cells partially overlap with the window. Consequently, the local variance grows with the window's surface area \cite{Fa14,Ki19a, Ki19b,Ch21c}.
This heuristic argument has recently been rigorously confirmed by proving that the weighted point process is hyperuniform for a large class of point processes with exponentially fast decay of correlations; see Example 8.4 in Ref. \cite{Kl25b}.
While Farago \textit{et al.} \cite{Fa14} and Chieco and Durian \cite{Ch21c} measured some of the weight-averaged local variances, spectral densities, and correlation functions, they did not provide mathematical relationships for these three quantities (as laid out in Secs. \ref{auto-spec} and \ref{variance}) or examine the conversion from antihyperuniform unweighted point configurations to hyperuniform weighted ones.
We also compute the weighted hyperuniformity order
metric defined by Eq. (\ref{eq:Lambda-exact}) of the different models to rank order them according to the degree to which they suppress large-scale fluctuations; see Appendix \ref{app:Lambda_f}.

\subsection{Exact results for 1D Poisson point configurations}
\label{app:1D-Poisson}

We begin by considering a 1D statistically homogeneous (stationary) Poisson point process (i.e., ideal gas in the grand canonical ensemble) at number density $\rho$ in the thermodynamic limit ($N\to\infty$). This represents a prototypical standard nonhyperuniform system in which the
structure factor is unity for all wave numbers, and the local number variance is exactly given by $\sigma_N^2(R)=2\rho R$ for all $R$.

The particle weights are the volumes (or the lengths) of the Voronoi cells, denoted by $v_i$.
We obtain exact results for the weight-averaged autocovariance function and spectral density. We show that the system is 
hyperuniform of class I in which the spectral density obeys the scaling ${\tilde \chi}_v(k) \sim k^4$. This result provides insights
concerning similar scalings for other 2D and 3D models described below.

The Voronoi length weights are deterministically given by the positions of the points, $v_i = (r_{i+1}-r_{i-1})/2$, where, without loss
of generality, we assume the point positions are arranged in ascending order, i.e.,
\begin{align}
 r_i < r_{i+1} \quad \text{for all }i\in\Z.
\end{align}

The probability density function of the volume of a typical cell is given by 
\begin{align} \label{eq:v-density-1D-Poisson}
 p(v) = 4\rho^2ve^{-2\rho v},
\end{align}
with $\langle v \rangle = \int_0^{\infty} p(v)vdv = 1/\rho$.
\begin{widetext}
\noindent We employ the generic two-particle probability density function $\rho_2(r,v_1,v_2)$ derived in Ref. \cite{Kla14} [see Eq. (A9) therein]:
\begin{align} \label{eq:rho2-1D-Poisson}
 \rho_2(r;v_1,v_2) 
 = \rho^2 a_1(r; v_1) \begin{cases}
 a_2(r; v_1,v_2)    & \text{if } v_1 < r/2 \\
 \frac{1}{r\rho+1} [r\rho a_3(r; v_1,v_2) + a_4(r; v_1,v_2)]   & \text{otherwise},
 \end{cases}
\end{align}
where we have used the auxiliary functions
\begin{align}
 a_1(r;v) &\equiv 
 \begin{cases}
  4v\rho^2e^{-2v\rho}    & \text{if }v<\frac{r}{2}\\
  2\rho(r\rho+1)e^{-2v\rho} & \text{otherwise}
 \end{cases}, 
 \label{eq:a1}\\
 a_2(r;v_1,v_2) &\equiv 
 \begin{cases}
  4v_2\rho^2{e}^{-2v_2\rho}                                  & \text{if }v_2 \le \frac{r}{2}-v_1 \\
 \frac{\rho{e}^{-2v_2\rho}}{2v_1}\left[4v_2(r-v_2)\rho-(r-2v_1)^2\rho+4(v_1+v_2)-2r\right]    & \text{else if } v_2 < \frac{r}{2} \\
  2\rho{e}^{-2v_2\rho} \left[1+\rho(r-v_1) \right]                     & \text{otherwise}
 \end{cases},
 \label{eq:a2}
\end{align}
\begin{align}
 a_3(r; v_1,v_2) &\equiv 
 \begin{cases}
 \frac{4v_2\rho{e}^{-2v_2\rho}}{r}((r-v_2)\rho+1)    & \text{if }v_2 \le \frac{r}{2} \\
 \rho{e}^{-2v_2\rho}(r\rho+2)    & \text{otherwise}
 \end{cases}, 
 \label{eq:a3}
 \\
 a_4(r;v_1,v_2) &\equiv 2\rho{e}^{r\rho}{e}^{-2v_2\rho} .
 \label{eq:a4}
\end{align}
\end{widetext}

The generalized total correlation function \eqref{h} of this model is obtained by using Eqs. \eqref{eq:v-density-1D-Poisson} and \eqref{eq:rho2-1D-Poisson}:
\begin{align} \label{eq:h-1D-Poisson}
 h(r;v_1,v_2) = \frac{\rho_2(r;v_1,v_2)}{\rho^2} - 16\rho^4v_1v_2e^{-2\rho (v_1+v_2)}
\end{align}
Inserting Eq. \eqref{eq:h-1D-Poisson} to Eq. \eqref{chi-r-1} yields an expression for the weight-averaged autocovariance function:
\begin{align} \label{eq:auto-1D-Poisson}
 \chi_v(r) = \frac{3\delta(r)}{2\rho} + \rho^2 h_v(r),
\end{align}
where $h_v(r)$ is the weight-averaged total correlation function exactly given by
\begin{align} \label{eq:hv-1D-Poisson}
h_v(r) =& \frac{1}{8 \rho^2} [(\rho r)^2 - 2(\rho r) - 6] e^{-\rho r},
\end{align}
and we have used the fact that $\expval{v^2}=3/(2\rho^2)$ from Eq. \eqref{eq:v-density-1D-Poisson}.
The function $h_v(r)$ is plotted in Fig. \ref{fig:1D-Poisson-h}(a).

Because all moments of the autocovariance are bounded, the corresponding
spectral density is analytic at the origin, i.e., its Taylor series involves only even powers of $k$.
Indeed, the exact expression for the spectral density is given by
\begin{align}  \label{eq:chivk-1D-Poisson}
 \tilde{\chi}_v(k) = \frac{( k/\rho)^4}{2 \rho} \frac{7+3(k/\rho)^2}{[1+( k/\rho)^2]^3}.
\end{align}
We see that for small $k$, the spectral density has the following scaling $\tilde{\chi}_v(k)\sim k^{\alpha_v}$ with
$\alpha_v=4$ and, hence, the change in the exponent in going from unweighted to weighted configurations is $\delta \alpha=4$, where $\delta \alpha$ is defined in Eq. (\ref{eq:Delta_alpha}).
Hence, the unweighted 1D Poisson point process has been converted 
to a weighted hyperuniform system of class I. 
The spectral density is plotted in Fig. \ref{fig:1D-Poisson-h}(b).

It is instructive to contrast these results with the work of Gabrielli {\it et al.} \cite{Ga08}, who showed that equal-volume tilings
of $\mathbb{R}^d$ yield unweighted disordered hyperuniform point configurations with a
 small-$k$ structure factor scaling $S(k) \sim k^4$ when the points are located at the centroids of each tile such that the shapes
and orientations of the tiles are short-ranged correlated. Thus, our present findings for 1D Poisson point configurations show that a scaling exponent of 4
can be achieved under more general conditions in which tiles (Voronoi cells in this case) have unequal volumes, the points
are not located at the tile centroids, but are weighted by their volumes. The same scaling exponent of 4 will characterize
Poisson point configurations in higher dimensions for reasons similar to those
put forth in Ref. \cite{Ki19b}.

The corresponding local variance $\weighted{\sigma^2}{\vect{f}}{R}$ for the weighted 1D Poisson point configurations can be obtained by substitution of 
relation (\ref{eq:chivk-1D-Poisson}) into (\ref{var-Fourier}) to yield the exact result
\begin{equation}\label{eq:sigma-1D-Poisson}
\weighted{\sigma^2}{\vect{f}}{R} = \frac{1}{\rho^2} \qty{ 1+ \exp(-2\rho R)\left[(\rho R)^2 +\rho R -1\right]}.
\end{equation}
We see that the local variance is bounded for all $R$ and tends to the constant $1/\rho^2$ for large $R$,
again implying hyperuniformity of the weighted 1D point configuration.
The local variance is plotted in Fig. \ref{fig:1D-Poisson-h}(c).
Its corresponding hyperuniformity order metric is exactly equal to the unity, i.e., $\overline{\Lambda}_\vect{f}=1$; see Table \ref{tab:Lambda}.
In contrast, the local number variance of the unweighted Poisson point configuration grows linearly with the window volume, i.e., $\sigma_N^2(R)=2\rho R$.

\begin{figure*}[bthp]
\includegraphics[width=0.9\textwidth]{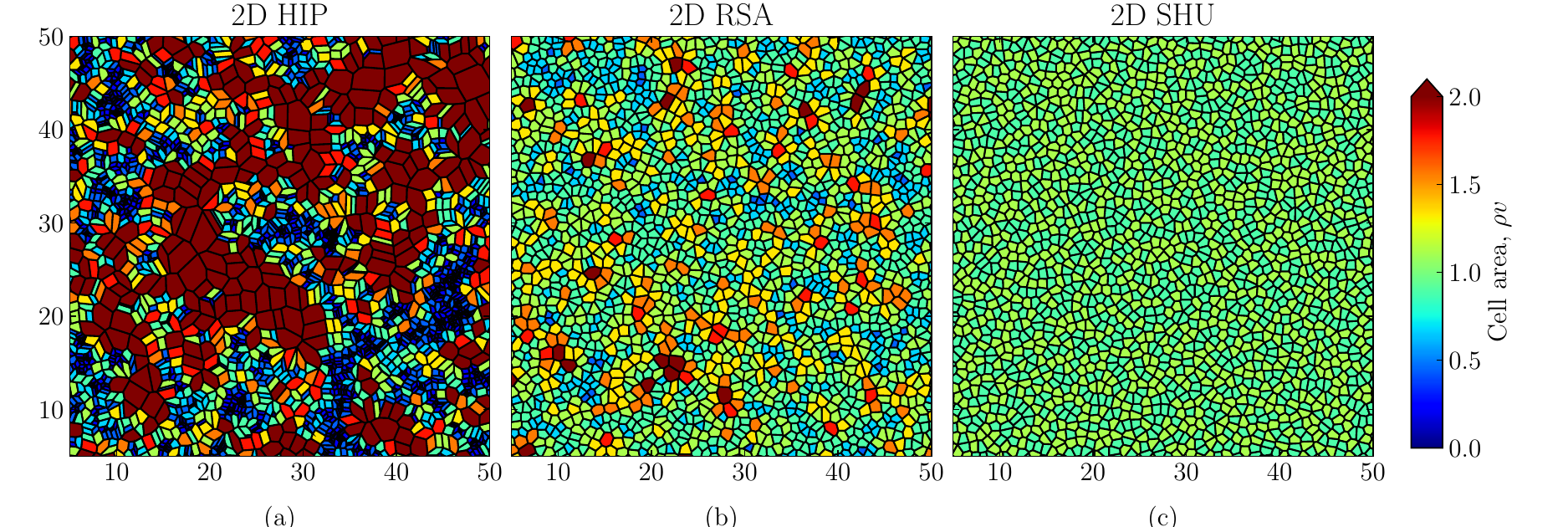}
\caption{ Representative images of 2D weighted point configurations derived from (a) HIP, (b) RSA (with a packing fraction $\phi=0.30$), and (c) SHU (with a
stealthiness parameter $\chi=0.35$) using the dimensionless Voronoi cell-areas $\rho v$ as weights.
Each Voronoi cell is colored according to the area of its Voronoi cell.
 \label{fig:2D_volumes_rep}
}
\end{figure*}

\subsection{2D models}
\label{s:volumes_2D}

Here, we study three different models of 2D weighted point configurations derived from 2D models: the hyperfluctuating or antihyperuniform \cite{To18a} hyperplane intersection process (HIP), the nonhyperuniform random sequential addition (RSA) process, and stealthy hyperuniform (SHU) point configurations. 
The points of the HIP are defined as the vertices (i.e., intersections of $d$ hyperplanes) of a 
Poisson hyperplane process, that is, of randomly and independently distributed hyperplanes; see Refs.~\cite{Schn08, Chi13} for details.

\begin{figure*}[bthp]
 \subfloat[]{\includegraphics[width=0.4\textwidth]{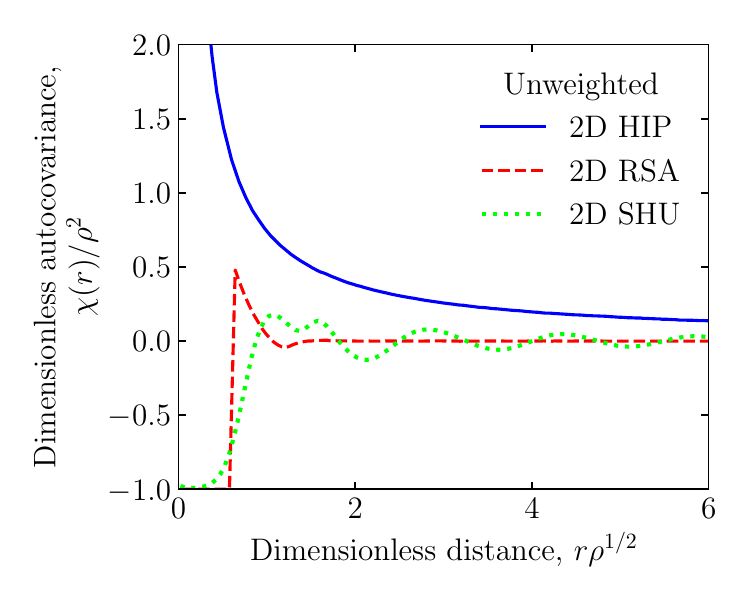}
 \includegraphics[width=0.4\textwidth]{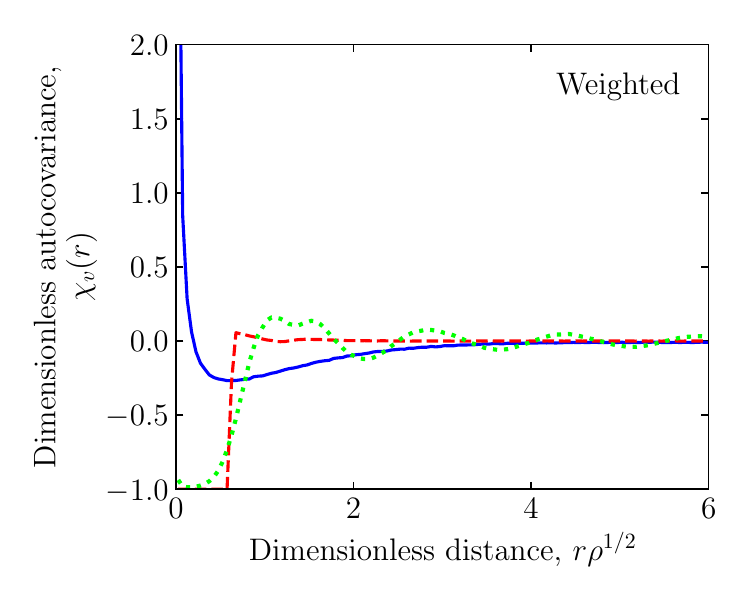}}

 \subfloat[]{\includegraphics[width=0.4\textwidth]{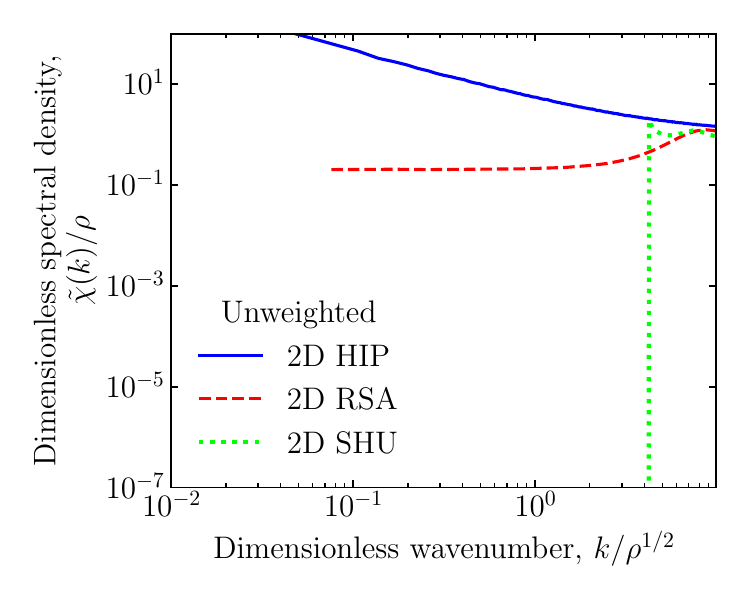}
 \includegraphics[width=0.4\textwidth]{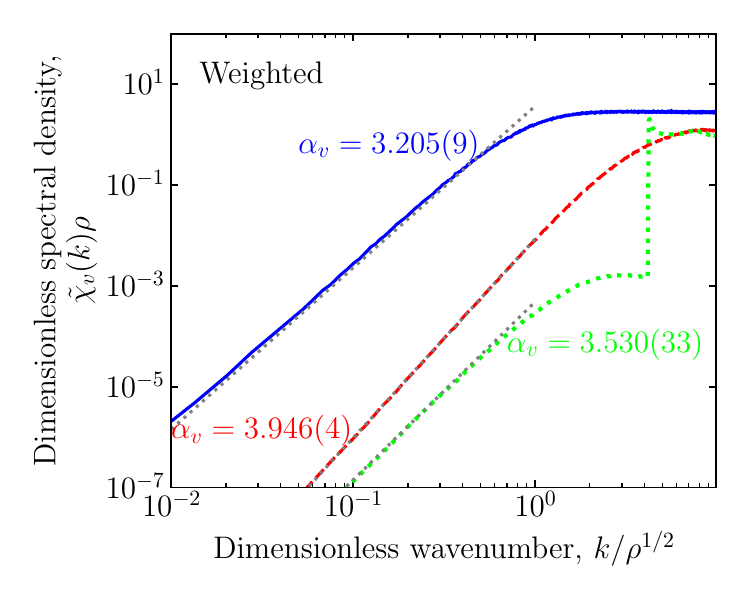}}

 \subfloat[]{
 \includegraphics[width=0.4\textwidth]{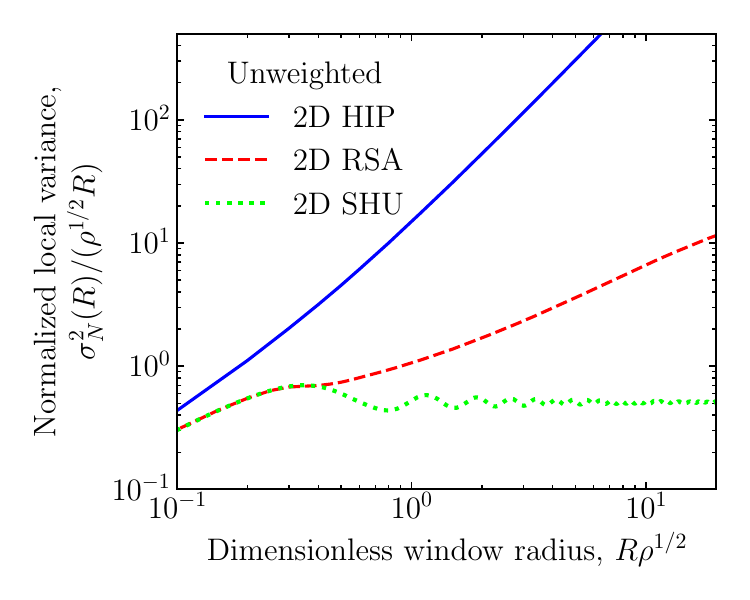}
 \includegraphics[width=0.4\textwidth]{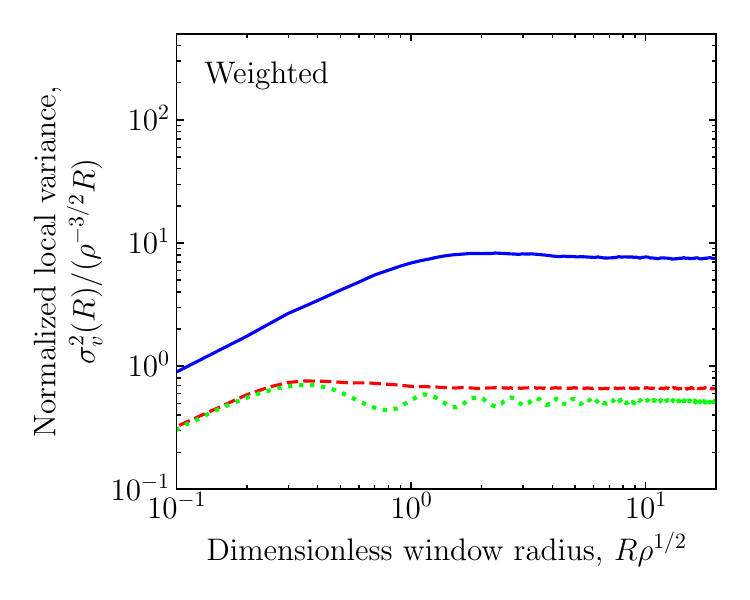}
 }
\caption{
Comparisons of the pair statistics for (Left) 2D unweighted point
patterns and (Right) the weighted ones derived from them by using
Voronoi cell areas $v$ as the particle weights. We consider HIP (with $\expval{N}=6.4\times 10^5$), RSA
(with $\phi=0.30$ and $N=9\times 10^4$) and SHU (with $\chi=0.35$ and $N=2\times 10^4$) models in each panel.
 (a) Dimensionless autocovariance function as a function of the
dimensionless distance $r\rho^{1/2}$.
 We show $\rho^{-2}\fn{\chi}{r}$ and $\fn{\chi_{_v}}{r}$ for the unweighted and
weighted cases, respectively.
 (b) Log-log plot of the dimensionless spectral density as a
function of the dimensionless wave number $k\rho^{-1/2}$.
 We show $\rho^{-1}\fn{\tilde{\chi}}{k}= S(k)$ and $\rho
\fn{\tilde{\chi}_{_v}}{r}$ for the unweighted and weighted cases,
respectively.
 (c) Log-log plot of the normalized local variance as a function of
the dimensionless window radius $R\rho^{1/2}$.
 We show $\fn{\sigma_N^2}{R}/(\rho^{1/2} R)$ and $\rho^2
\fn{\sigma_v^2}{R}/(\rho^{1/2} R)$ for the unweighted and weighted
cases, respectively.
  \label{fig:2D_volumes_auto}
 } 

\end{figure*}

The local number variance of the HIP 
 scales faster than the volume of the observation window, since the structure factor diverges in the limit $k \to 0$~\cite{He06, Kl19a}.
The system size of this model is characterized by the mean particle number $\expval{N}$ because each configuration has a different particle number $N$,
even if the deviations is small.
We consider large HIP configurations in a range of system sizes $\expval{N}= 4.9 \times 10^{3} - 6.4\times 10^5$, and present in Fig. \ref{fig:2D_volumes_auto} the autocovariance function, the spectral density (along with the associated hyperuniformity exponent), and the local variance for the largest systems.
Estimates of the exponents for other system sizes are listed in Table \ref{tab:hip} in Appendix \ref{app:exponents}.

The RSA process is a time-dependent (nonequilibrium) packing procedure that generates disordered sphere packings in $\mathbb{R}^d$ \cite{Re63,Wi66,Fe80,Co88,To06a,Zh13b}. 
Starting with an empty but large volume in $\mathbb{R}^d$, the RSA process is produced by randomly, irreversibly, and sequentially placing nonoverlapping spheres into the volume. 
This procedure can stop at a range of packing fractions $\phi$ up to the maximal saturation value $\phi_s$ \cite{Zh17a}, which 
in two dimensions is approximately equal to 0.55 \cite{Re63,Fe80,Co88,To06d,Zh13b}.
It is known that the RSA process in $\mathbb{R}^d$ is characterized by a pair correlation function that decays to unity
super-exponentially fast, even at saturation \cite{Bo94}.
We consider RSA configurations with $\phi=0.30$ and $N= 4\times 10^2 - 9\times 10^4$ , and report in  Fig. \ref{fig:2D_volumes_auto} the autocovariance function, the spectral density (along with the associated hyperuniformity exponent), and the local variance for the largest systems.
Estimates of the exponents for other system sizes are listed in Table \ref{tab:rsa} in Appendix \ref{app:exponents}.

The SHU configurations are defined by a structure factor that vanishes in a spherical region around the origin in Fourier space, 
i.e., $S({\bf k})=0$ for $0<|{\bf k}|\leq K$. They are class I hyperuniform.
A powerful procedure that enables one to generate high-fidelity disordered stealthy
hyperuniform point configurations is the {\it collective-coordinate} optimization technique
\cite{Fa91,Uc04b,Uc06b,Ba08,To15,Zh15a}. This optimization methodology involves finding the highly degenerate ground states in the disordered
regime of a class of bounded pair potentials with compact support in Fourier space, which
are stealthy and hyperuniform by construction. 
The {\it stealthiness parameter} $\chi$ is a dimensionless measure of the ratio of constrained degrees of freedom (i.e., wave vectors contained within the cutoff
wave number $K$) to the total degrees of freedom (approximately $d N$) in such an optimization procedure. 
A point configuration with a small value of $\chi$ (relatively unconstrained) is disordered, and as $\chi$ increases, the short-range order increases within a disordered regime ($ \chi < 1/2$ for $d=2$ and $d=3$) \cite{To15}. 
We study disordered SHU configurations with $\chi=0.35$ and $N=4\times 10^3-2\times 10^4$, and report in Fig. \ref{fig:2D_volumes_auto} the autocovariance function, the spectral density (along with the associated hyperuniformity exponent), and the local variance  for the largest systems.
Estimates of the exponents for other system sizes are listed in Table \ref{tab:shu} in Appendix \ref{app:exponents}.

Figure \ref{fig:2D_volumes_rep} shows representative images of the weighted point configurations derived from the three models mentioned above.
For visualization purposes, we color each Voronoi cell according to its dimensionless area $\rho v$.
It is seen that the Voronoi cell areas become more narrowly distributed in going from 
the unweighted antihyperuniform HIP, nonhyperuniform RSA, and stealthy hyperuniform point configurations.
The cell-area probability density functions of these three models are plotted in Appendix \ref{prob}.
At first glance, one might surmise that the weighted antihyperuniform HIP cannot be hyperuniform, but
this is counterintuitively not the case, as we show immediately below.

Figure \ref{fig:2D_volumes_auto} compares the pair statistics of 2D
unweighted point configurations and the weighted ones derived from them.
In converting the 2D point configurations to the weighted ones, the
changes in the autocovariance function are the least for SHU and the greatest for HIP; see Fig. \ref{fig:2D_volumes_auto}(a), since the Voronoi cell areas in the former are narrowly distributed around their mean value in contrast to the latter, which has a broad distribution.
Importantly, we find that all unweighted configurations, including the HIP,
become class I in their weighted counterparts.
Figure \ref{fig:2D_volumes_auto}(b) shows that regardless of the degree of long-wavelength fluctuations in the unweighted point configurations, the weighted ones have a hyperuniformity exponent $\alpha_{\bf f}$ that ranges between 3 and 4, and, hence, are all of class I.
The class I hyperuniformity of all three models is also reflected in the fact that the local variance scales as $\fn{\sigma_v^2}{R}\sim R$ for large $R$
 as shown in Fig. \ref{fig:2D_volumes_auto}(c).

Specifically, in the case of the RSA model, we find that the hyperuniformity exponent is $\alpha_v=3.946\pm 0.004$ for the largest system size analyzed here ($N=9\times 10^4$),
and our finite-$N$ analysis suggests that $\alpha_v$ converges to 4 from below in the thermodynamic limit and so $\delta\alpha=4$.
Thus, the weighted RSA process has the same scaling behavior as the corresponding Poisson point process.
Apparently, the superexponential decay of pair correlations in the former imbues it with the same behavior in the latter, which is completely spatially uncorrelated. Consequently, we can conclude that the weighted spectral density
for the RSA model is analytic at the origin, implying that all of the moments of its autocovariance function
are bounded.

\begin{figure*}[t]
\subfloat[]{\includegraphics[width=0.32\textwidth]{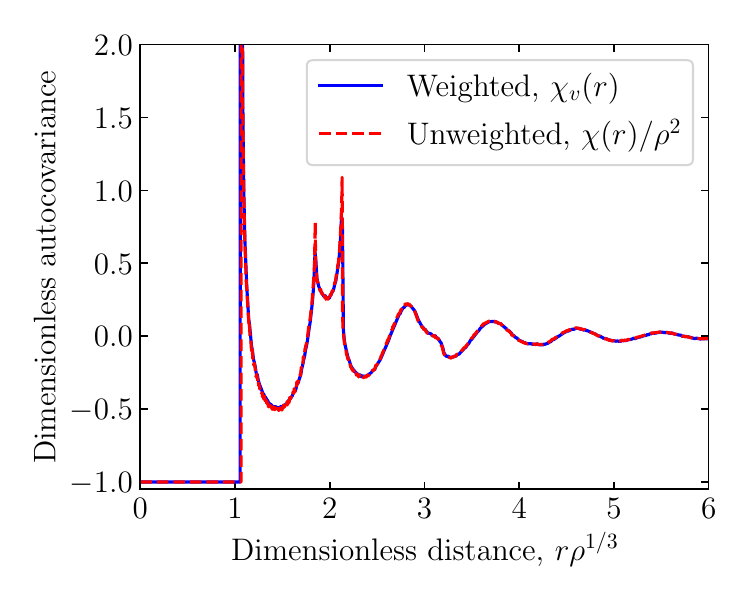}}
\subfloat[]{\includegraphics[width=0.32\textwidth]{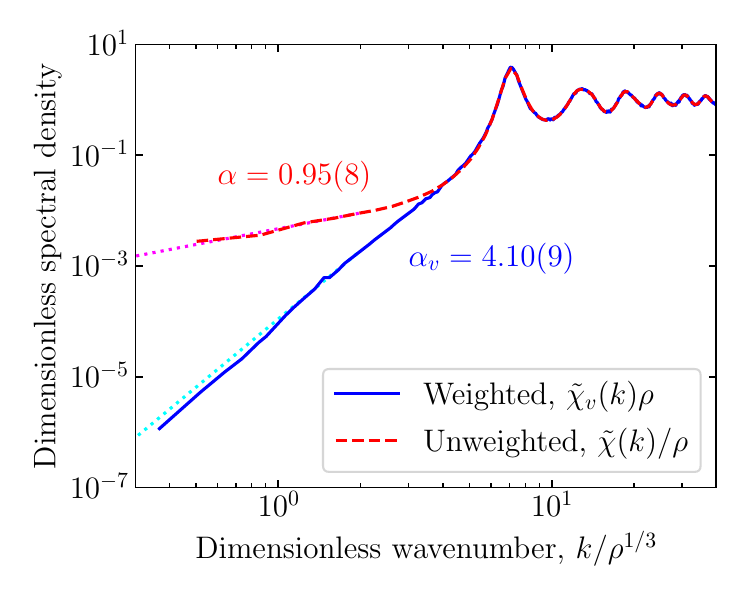}}
\subfloat[]{
\includegraphics[width=0.32\textwidth]{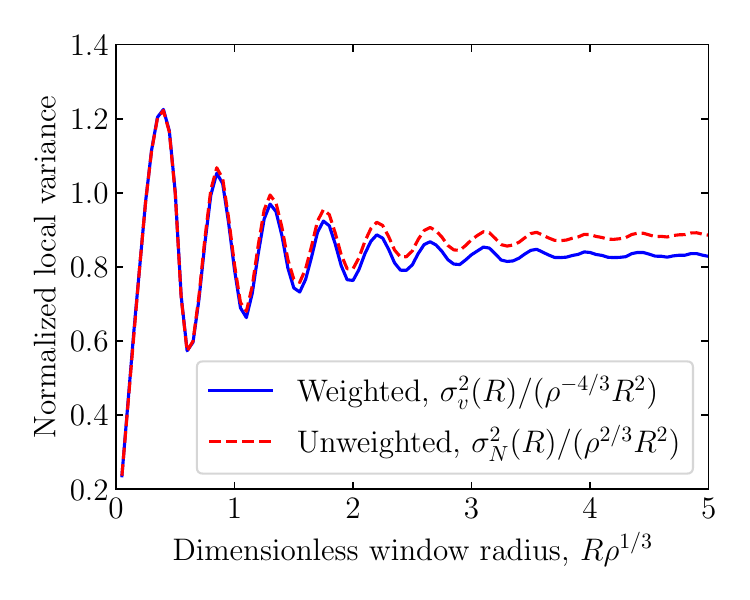}}
\caption{Comparisons of 3D MRJ packings and the associated weighted point configurations, using Voronoi-cell volume $v$ as a weight $f_j$ of particle $j$, and unweighted ones.
 (a) Dimensionless autocovariance function. For the weighted and unweighted cases, the $y$-axes become $\fn{\chi_{_v}}{r}$ and $\fn{\chi}{r} / \rho^{2}$, respectively.
 (b) Log-log plot of the dimensionless spectral density. For the weighted and unweighted cases, the y-axes become $ \fn{\tilde{\chi}_{_v}}{k} \rho$ and $ \fn{\tilde{\chi}}{k} / \rho$, respectively.
 (c) The dimensionless local variance normalized by $(\rho^{1/3}R)^2$. For the weighted and unweighted cases, the $y$-axes become $\rho^2 \sigma_{_v}^2(R)/(\rho^{-4/3}R^2)$ and $\sigma_N^2(R)/(\rho^{2/3}R^2)$
 Configurations of 3D MRJ packings were taken from Ref. \cite{Ma23}.
 \label{fig:3D_MRJ_stat}}
\end{figure*}

For the largest SHU system analyzed here ($\chi=0.35$, $N=2\times 10^4$), we find that $\alpha_v = 3.530 \pm 0.033$, which is expected to converge to $3.5$ 
in the thermodynamic limit and, hence, $\delta\alpha= -\infty$. This result
implies that its weight-averaged spectral density is nonanalytic at the origin, and, hence, the corresponding autocovariance function decays with the power-law
tail $1/r^{5.5}$. For the HIP model, we find an exponent $\alpha_v=3.205\pm 0.009$ as the system size increases from $\expval{N}=70^2$ to $800^2$.
From this trend, the exponent is expected to converge to $3$ in the thermodynamic limit, meaning that $\delta\alpha=3-(-1)=4.0$.
Thus, the weighted spectral density for HIP is also nonanalytic at the origin with an autocovariance function decaying with the power-law
tail $1/r^{5}$. Why are the exponents for the SHU and HIP models nonintegers or smaller than the exponent of 4
that applies to the Poisson and RSA models? This may be explained using the aforementioned results
of Gabrielli, Joyce, and Torquato \cite{Ga08}, who also showed that the scaling exponent can be smaller than 4
due to the long-range correlations in the first moment of mass distribution of Voronoi cells with respect to the particle centers, denoted by $v_i \Delta \vect{x}_i$, where $\Delta \vect{x}_i$ is the displacement vector from particle $i$ to the
centroid of its Voronoi cell \cite{Ga08}\footnote{The Taylor expansion of the structure factor performed by Gabrielli {\it et al.} \cite{Ga08} can be employed in the weighted point configurations with some minor revisions. In Eq. (20) therein, HIP has an additional term $\Tr[\tilde{g}^{1,1}(\vect{k})]$ that accounts for the differences in particle centers and cell centroids (i.e., $\Delta \vect{x}_i \neq \vect{0}$).
Due to the long-range correlations in the first moments $v_i \Delta \vect{x}_i$, we have $\Tr[\tilde{g}^{1,1}(\vect{k})]\sim \abs{\vect{k}}$ for small wave numbers. Consequently, the leading-order behavior of its weight-averaged spectral density becomes $\tilde{\chi}_v(k) \propto k^2 \Tr[\tilde{g}^{1,1}(\vect{k})] /2 \sim k^3$.}.
Both the SHU and HIP models are characterized by Voronoi cells with long-range correlations.


Table \ref{tab:Lambda} lists
the hyperuniformity order metric $\overline{\Lambda}_v$ for each of these class I models.  The HIP model shows the highest degree of large-scale fluctuations, followed by the RSA, and the SHU shows the least. Interestingly, this ranking is consistent with those of their unweighted 
counterparts, even if only one of them (SHU)
is a hyperuniform case.

\subsection{3D MRJ packings}  
\label{s:3D_MRJ_volumes}

Here we consider Voronoi-cell volume weighted configurations of 
3D maximally random jammed (MRJ) sphere packings \cite{Do05d,To10e,Ch12,Ma23}, which are prototypical glasses, since they are maximally disordered, perfectly rigid, and perfectly nonergodic.
 Such disordered jammed packings, which are hyperuniform \cite{Do05d,Ma23} are best produced by using the linear programming packing algorithm of Torquato and Jiao \cite{To10e}. Recently, it has been shown that
such states, to an excellent approximation, can be generated as disordered SHU ground states in the zero-$\chi$ limit \cite{To25a}.

To characterize the properties of such nonequilibrium jammed packings, Klatt and Torquato \cite{Kla14} studied certain correlation functions of 
the Minkowski functionals (e.g., volume, surface area, and integrated mean curvature) of their Voronoi cells. 
Importantly, in the case of the Voronoi-cell volumes, they studied the correlation function $C_{00}({\bf r})$, which is distinctly different from $\chi_v({\bf r})$ \footnote{These two functions are related as follows: $ 
 C_{00}({\bf r}_1, {\bf r}_2) = \frac{\chi_{_v}({\bf r}_1, {\bf r}_2)}{\sigma_{v({\bf r}_1 | {\bf r}_2)} \sigma_{v({\bf r}_2 | {\bf r}_1)}},
 $
  where $\sigma_{v({\bf r}_1 | {\bf r}_2)}$ is the standard deviation of the Voronoi-cell volume at ${\bf r}_1$ given that there is another point at ${\bf r}_2$.}.
They also studied the probability density function $g_{vv}(\vb{r};v,v')$, which is precisely the same as our generalized pair correlation function $g_{2,v}(\vb{r};v,v')$ that is not averaged over weights; see Eq. \eqref{eq:generalized-g2} for any type of weights.
 Importantly, the function $C_{00}(\vb{r})$ does not reflect the split-second peak [unlike the function $g_{vv}(\vb{r};v,v')$] and the power-law singularity for near contacts
exhibited by the standard pair correlation function $g_2(\vb{r})$ of 3D MRJ packings \cite{To18b}.
 
In Fig. \ref{fig:3D_MRJ_stat}(a), we compute the weighted autocovariance function $\chi_{_v}({\bf r}_1, {\bf r}_2)$, which heretofore has not been determined.
The corresponding autocovariance for the unweighted configurations is also shown.
In the present work, we also determine
both the generalized spectral density and local variance for MRJ sphere packings in which $f_j=v_j$,
which was not done in Ref. \cite{Kla14}.

It has been reported that 3D MRJ sphere packings are class II hyperuniform \cite{Do05d, Ma23}, meaning that the structure factor (or, equivalently, the unweighted spectral density) exhibits a power-law scaling $S(k)\sim k$ for small $k$; see Fig. \ref{fig:3D_MRJ_stat}(b). 
 This also implies that the local number variance (or, equivalently, the unweighted local variance) exhibits the large-$R$ asymptotic behavior, i.e., $\sigma_N^2 (R) \sim R^2 \ln(R)$ for large $R$; see Fig. \ref{fig:3D_MRJ_stat}(c).

 3D MRJ Voronoi-cell-volume-weighted configurations suppress long-wavelength fluctuations to a greater extent than the unweighted ones.
 Specifically, they now become class I hyperuniform with $\tilde{\chi}_{v}(k)\sim k^\alpha_{v}$ for small $k$ with $\alpha_v=4.1\pm 0.09$ (see Fig. \ref{fig:3D_MRJ_stat}(a)) and $\sigma_v^2(R)\sim R^2$ for large $R$ (see Fig. \ref{fig:3D_MRJ_stat}(c)). 
We expect that $\alpha_v=4$ in the thermodynamic limit and, hence, $\delta \alpha=3$.
 We note that these results are consistent with spectral density and local variances of the 3D sphere packings with polydispersity
in size that were created from 3D point configurations by decorating each point with a sphere whose volume is proportional to the associated Voronoi-cell volume \cite{Ki19a,Ki19b}.
Table \ref{tab:Lambda} lists the value
the hyperuniformity order metric $\overline{\Lambda}_v$
for this 3D model.

\section{Disordered 2D Charged Systems (Ionic Liquids) Mapped From Excess Side Numbers of Voronoi Cells}
\label{charges}

\begin{figure*}[ht]
 \centering
 \includegraphics[width=0.8\textwidth]{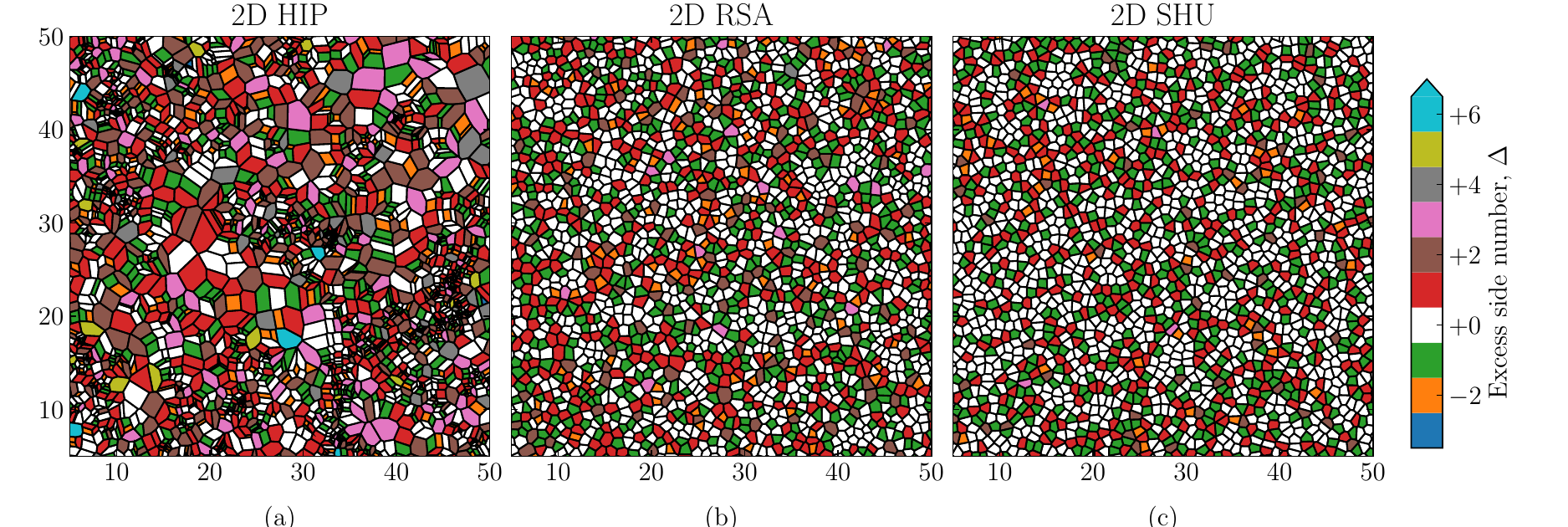}
 \caption{Representative images of 2D weighted point configurations derived from (a) HIP, (b) RSA ($\phi=0.30$), and (c) SHU ($\chi=0.35$), using the excess side numbers $\Delta\equiv s-6$ as weights. 
 Each Voronoi cell is colored according to the values of $\Delta$. 
The cell-area probability distributions of these three models are plotted in Appendix \ref{prob}.
}
  \label{fig:2D-excess-representative}
\end{figure*}

Charged many-particle systems, such as classical Coulomb systems \cite{Ma80,Leb83} and ionic liquids \cite{Le97,An12b,Le19c},
are excellent examples of configurations with scalar weights, namely, the electric charges carried by the particles.
Various types of plasmas in equilibrium, free charged particles of one sign in a rigid background of opposite charge for overall
charge neutrality, have been shown to be hyperuniform of class I 
in their unweighted particle densities \cite{Dy70,Fo10,Lo17,Lo18a,To18a}.
It has been found that the structure factor scaling exponent $\alpha$ is equal to 2 for $d$-dimensional equilibrium systems (Gibbs ensembles) subject to Coulombic
interactions in both the one-component \cite{Ha73,Ja81} and two-component plasmas \cite{Lo17,Lo18a,To18a}.
In cases when the $(d+1)$-dimensional Coulombic interactions are in $\mathbb{R}^d$, the scaling exponent
is equal to 1 \cite{Dy70,To18a}. Torquato \cite{To18a} showed that $\alpha$ can be prescribed
 by a generalized Coulombic pair potential $v(r)$ of ``like-charged'' particles in which the
 long-ranged asymptotic form of $v(r)$ is given by
\begin{equation}
 v(r)\sim 
 \begin{cases}
 r^{-(d-\alpha)}, & d \ne \alpha\\
 -\ln(r), & d = \alpha,
 \end{cases}
  \label{v_long_hu}
\end{equation}
For $\alpha = 2$, the long-ranged parts of the potentials are given by the Coulomb interactions in the respective dimensions.
For all of these plasmas, the rigid background and overall charge neutrality ensure 
hyperuniformity in the charge density fluctuations of the free charged particles.
It is well-established that charge fluctuations in general equilibrium Coulomb systems (including nonplasmas,
such as molten salts, ionic liquids)
are provably hyperuniform of class I under mild clustering conditions and the decay of pair correlations \cite{Ma80,Leb83,Le97}.
However, the scaling exponent $\alpha$ was not considered in these studies, since
this value, which must be greater than unity, is system dependent.

Here we introduce the concept of weights derived from the ``excess'' number of sides of Voronoi cells (defined below) associated with 2D statistically homogeneous disordered point configurations, which can be mapped to certain charged systems.
The average number of sides of Voronoi cells in such disordered systems is 6 due to Euler's formula \cite{To02a}. 
Therefore, the number of sides $s$ of a cell in excess of this mean
value, which we denote by $\Delta \equiv s-6$ and call the {\it excess side number}, will take integer values, either positive, negative or zero.
The excess side number $\Delta$ can be viewed as charges whose values, concentrations
and spatial distributions are imposed by the Voronoi network 
such that the entire charged system possesses overall charge neutrality, since the mean of $\Delta$ is exactly zero due to Euler's formula.
Moreover, such static disordered charged systems can be regarded as ionic liquids that are {\it out of equilibrium} because of the  geometric mapping from static
point configurations.
Generally, the positions of the charges are correlated, since the excess side numbers are constrained by
the Voronoi network such that charge fluctuations are hyperuniform of class I due to  overall charge neutrality and effective {\it screening}, as we discuss below. 

\begin{figure*}[ht]
 \centering
 {\includegraphics[width=0.32\textwidth]{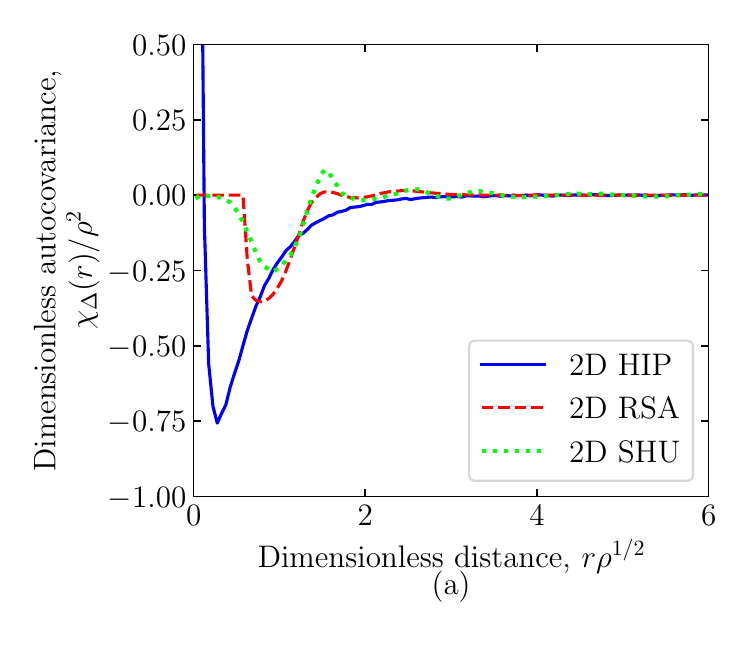}}
 {\includegraphics[width=0.32\textwidth]{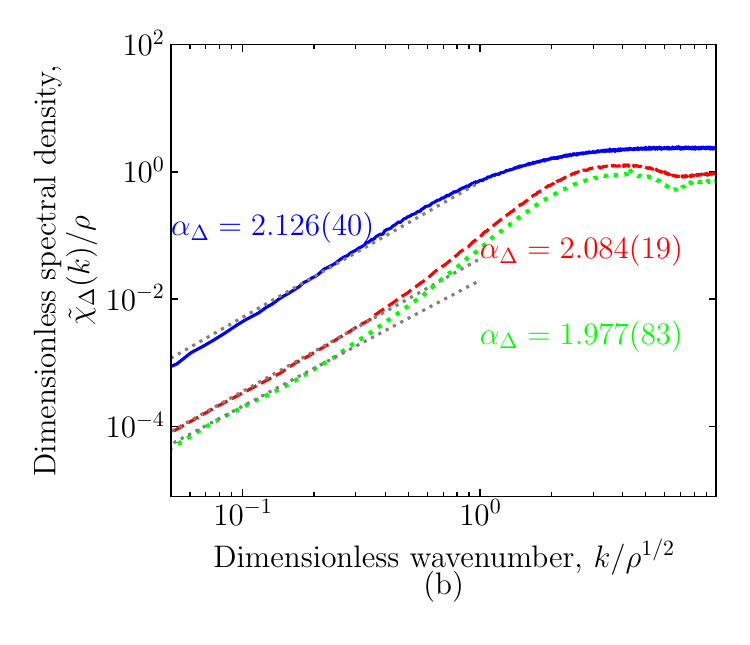}}
 {\includegraphics[width=0.32\textwidth]{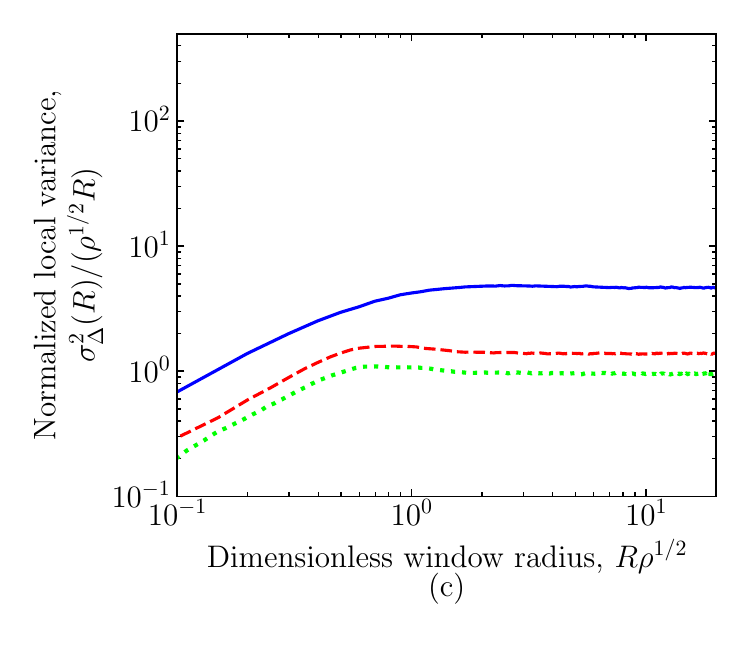}}
 \caption{Pair statistics and local variances of 2D weighted point configurations derived from HIP, RSA, and SHU, using the excess side numbers $\Delta$. 
 (a) dimensionless autocovariance function $\fn{\chi_{_\Delta}}{r} / \rho^{-2}$ as a function of the dimensionless distance $r\rho^{1/2}$.
 (b) Log-log plot of the dimensionless spectral density $\fn{\tilde{\chi}_{_\Delta}}{k} / \rho $ as a function of the dimensionless wave number $k/\rho^{1/2}$.
 (c) Log-log plot of the normalized local variance $\fn{\sigma_\Delta^2}{R}/(\rho^{1/2}R)$ as a function of the dimensionless window radius $R\rho^{1/2}$.
 }
  \label{fig:2D-excess-pair}
\end{figure*}

In particular, we study the weight $\Delta$ for the same 2D HIP, RSA and SHU models examined in Sec. \ref{volume} at the 
largest sample sizes.
Representative images of these weighted configurations are shown in Fig. \ref{fig:2D-excess-representative}.
The first two panels of Fig. \ref{fig:2D-excess-pair} show plots of the pair statistics of the excess side number weights (charges).
For all the models, the weight-averaged autocovariance function, $\chi_{_\Delta}(r)$, has both positive and negative correlations,
but they are short-ranged, i.e., it rapidly tends to zero for distances larger than $r \rho^{1/2}=3$. This 
implies that all of its moments exist and, hence, the associated spectral density $\fn{\tilde{\chi}_{_\Delta}}{k}$ is analytic at the origin. The latter implies that the Taylor series expansion of $\fn{\tilde{\chi}_{_\Delta}}{k}$ about $k=0$ can only have even powers of $k$.
(Unlike HIP, $\chi_{_\Delta}(r)$ of RSA and SHU also converge to zero at short distances $r\rho^{1/2}<0.1$ due to the nonoverlap constraint on the particles; see Fig. \ref{fig:2D-excess-pair}(a).)
Importantly, we confirm that the weighted point configurations are class I hyperuniform, regardless of the degree of long-wavelength fluctuations in the unweighted ones,
as shown in the plot of the spectral density depicted in Fig. \ref{fig:2D-excess-pair}(b).
For the finite samples considered here, the hyperuniformity exponent $\alpha_\Delta$ is approximately equal to 2.
In particular, HIP, RSA, and SHU models have the exponents $\alpha_\Delta=2.126\pm0.040, 2.084\pm0.019$, and $1.977\pm 0.083$, respectively, at the largest system sizes considered here; see Appendix \ref{app:exponents} for estimates of the exponents for other system sizes.

\begin{figure*}[ht]
 \includegraphics[width=0.8\textwidth]{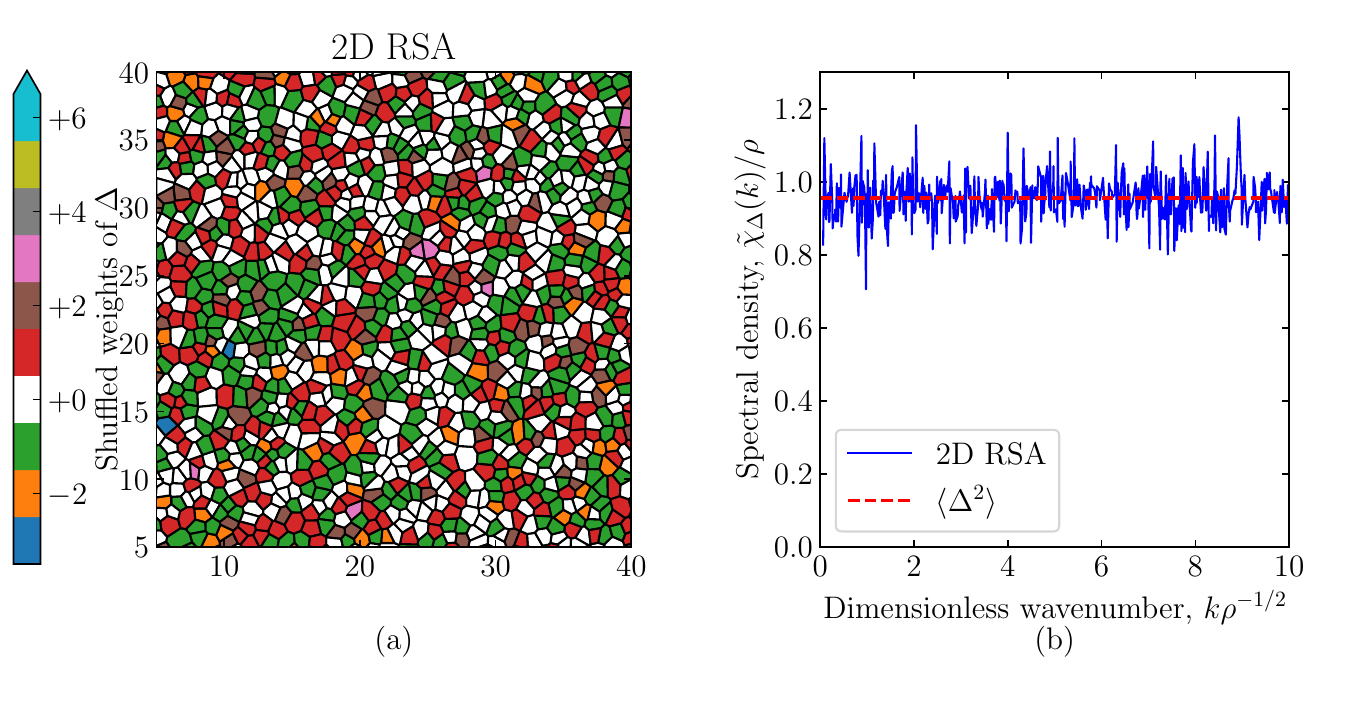}
 \caption{Two-dimensional RSA configuration with shuffled weights of $\Delta$. (a) The representative image of the weighted point pattern. The particle centers are identical to Fig. \ref{fig:2D-excess-representative}(b), but the weights are randomly shuffled without spatial correlation. 
 (b) Dimensionless spectral density $\fn{\tilde{\chi}_{_\Delta}}{k} / \rho$ of the corresponding weighted configurations as a function of the dimensionless wave number $k/\rho^{1/2}$. 
 The dashed line depicts $\expval{\Delta^2}$.}
  \label{fig:2D-RSA-shuffled}
\end{figure*}

Additionally, the exponent $\alpha_v$ commonly tends to converge to 2 from above as the system size increases.
In Fig. \ref{fig:2D-excess-pair}(c), we also confirm that the local variances of the weighted ones exhibit a common large-$R$ behavior with $\fn{\sigma_f^2}{R}\sim R$, indicating class I hyperuniformity. 
Note that while both Voronoi cell weights and excess side number weights produce class I 
hyperuniform systems, the large-scale fluctuations in the former with $\alpha_v=3.1-4$ are suppressed to a greater degree than in the latter with $\alpha_\Delta=2$.

Table \ref{tab:Lambda} summarizes the values of the hyperuniformity order metric $\overline{\Lambda}_\Delta$ for each of these class I models. The HIP model shows the highest degree of large-scale charge fluctuations, followed by the RSA model, and then the  SHU system, whose order metric is the smallest among all of
the three models. This ranking is consistent with what we observe for the models weighted by their Voronoi-cell volumes $v$; see also Fig. \ref{fig:2D_volumes_auto}(c).

Why are these charged nonequilibrium systems hyperuniform of class I?  As noted above, the charges are 
effectively screened due to the constraints imposed in the disordered Voronoi networks, leading to both
positive and negative correlations on relatively small length scales. Such correlations imply
strongly diminished charge fluctuations (due to effective neutrality) across length scales 
such that they are proportional to the window surface area for large windows. These screening phenomena are very reminiscent of
what occurs in equilibrium charged systems with long-range interactions \cite{Ma80,Leb83}, and hence appears to be common requirements to achieve class I hyperuniformity in equilibrium and nonequilibrium
charged systems alike.

To further reinforce the fact that the hyperuniformity of these 2D weighted point configurations comes from both 
effective screening and overall charge neutrality, we now show, by way of an example, that overall charge neutrality alone is not sufficient to 
obtain hyperuniform charge fluctuations. Specifically, beginning from the 2D RSA model with charges $\Delta$, we randomly interchange charges at the positions (i.e., without any spatial correlations)
 while maintaining overall charge neutrality. A portion of the resulting configuration of 2D RSA is shown in Fig. \ref{fig:2D-RSA-shuffled}, which is obtained from Fig. \ref{fig:2D-excess-representative}(b).
 Comparing the original configuration [Fig. \ref{fig:2D-excess-representative}(b)] with the randomly 
shuffled one [Fig. \ref{fig:2D-RSA-shuffled}(a)], one sees that the particles with the same charges 
(e.g., $\Delta=1$ colored in red or $\Delta=-1$ colored in green) tend to cluster more closely together,
indicating the lack of screening effects. Indeed, 
 the shuffled configuration is always nonhyperuniform because the charges are uncorrelated, i.e.,
the spectral density in the thermodynamic limit is simply equal to the positive constant $\rho\expval{\abs{\vect{f}}^2}$, as specified
by Eq. \eqref{uncorr}. The numerical evaluation of the spectral density, shown in Fig. \ref{fig:2D-RSA-shuffled}(b), is 
consistent with this theoretical result.

\section{Conclusions and Discussion}

A strong motivation for the present work is the fact
the large-scale fluctuations of weighted
many-particle systems have profound implications
for their physical properties,   which we discuss below, and
has already has been shown for
the unweighted counterparts, i.e., number fluctuations \cite{To18a}.
In this work, we considered particle weights that can be (complex-valued) scalars or vectors. 
We derived the generalized weighted pair correlation functions and
autocovariance functions as well as their corresponding  spectral
functions.   We then obtained formulas
for the weighted local variance, which are given in terms
of certain integrals involving either the aforementioned direct-space or Fourier-space pair statistics.  

Using this framework, we analyzed large-scale fluctuations in weighted systems, including bond-orientationally ordered phases, dipolar liquid water, Voronoi-cell volumes, and certain ionic liquids in different Euclidean space dimensions ($d=1,2$ and 3).
Reexamination of simulation data for liquid water led us to conclude that it is typically nonhyperuniform
with respect to its dipole moments (dipole variance growing in proportion to the window volume). We showed that it is not necessarily true that a hyperuniform or nonhyperuniform unweighted system will maintain the same behavior when weighted. In fact, we specifically demonstrate that unweighted systems can exhibit antihyperuniformity (i.e., weighted fluctuations that grow faster than the window volume) with respect to nematic, tetratic, and hexatic bond-orientational order.
Specifically, for the bond-orientationally ordered phases considered here, $\alpha_{\psi_n}$ ranges between $-2$ and $-1$.
For the 2D solid phase, we estimated $\alpha_{\psi_6}\approx -0.20$ for the equilibrium hard disks taken from Ref. \cite{Ber11}, and $\alpha_{\psi_6}\approx-0.90$ for ultradense SHU packings \cite{Ki25a}.
Conversely, we established that even when the underlying point configurations are nonhyperuniform or even antihyperuniform (i.e., the diametric opposite of hyperuniform), the corresponding weighted configurations can still exhibit hyperuniformity. We demonstrated that this scenario applies in cases where the weights are the Voronoi-cell volumes of particles across different spatial dimensions, as well as in systems of charged particles in certain 2D nonequilibrium ionic liquids, mapped from weights that are excess side numbers $\Delta$. 
We found that $\alpha_v$ typically ranges from $3$ to $4$, while our results for $\alpha_\Delta$ strongly suggest that it is equal to  2.

For the cases of weighted models studied here, there are many interesting open questions. For instance, how robust is the exponent of $\alpha_{\Delta}=2$ for the excess side numbers
beyond the 2D models considered here? Are there class I systems for 
which $\alpha_{\Delta}$ is greater than two? Are there class II 
structures, i.e., those with $\alpha_{\Delta}=1$?

Although such systems are ionic liquids out of equilibrium, it would be nonetheless interesting to see if one could still
extract an {\it effective} screening length.
One possible way of answering this question is to use the techniques
from Ref.~\cite{To22d} that enable a precise determination of effective pair interactions from pair statistics of many-body systems out of equilibrium.
For the problem at hand, one would assume
a set of basis functions that includes Yukawa interactions from which
an effective screening length would be extracted.
Answers to these questions would shed light on the correlations required to manipulate the large-scale charge fluctuations.
We observed that the 2D bond order in nematic/tetratic/hexatic phases and solid phases consistently led to antihyperuniform states.
Thus, in future work, it would be interesting to explore the 
possible existence of disordered hyperuniform systems that preserve
hyperuniformity when weighted by $\psi_n$.

Of course, the particular systems that were studied in the present work represent a small subset of the diversity of weighted many-particle systems that can now be analyzed using our theoretical formalism to quantify the nature of their large-scale fluctuations   and corresponding physical properties.
For example, homopolymer fluids in which the weight is the bond vector $\vect{l}_i$ of molecule $i$ can be examined using the correlation function $\expval{\vect{l}_1\cdot \vect{l}_2}$ and corresponding spectral density, which are essential for understanding their structural and dynamical properties \cite{Fl69,Wi10,Se10}.
In the context of nematic liquid crystals, the local deviation of the director field $\delta \vect{n}(\vect{r})\equiv \vect{n}(\vect{r})-\expval{\vect{n}(\vect{r})}$ and corresponding pair correlation functions $g_{\alpha\beta}(\vect{r})\equiv \expval{\delta n_{\alpha}(\vect{0})\delta n_{\beta}(\vect{r})}$ are key in determining their elastic properties and light scattering intensity at long wavelengths \cite{Sh94,De95,Ak04}.
Additionally, in statistically anisotropic systems with director weights, it would be interesting to ascertain whether they possess directional hyperuniformity, i.e., hyperuniformity that depends
on the direction in which the wavevector-dependent
spectral density tends to zero  \cite{To16a}.
Particle velocities in fluidized beds, sedimenting suspensions, and granular flows \cite{Lad96,Cow00,Ha15,Yu20} are natural weights
to consider under the hyperuniformity lens. Importantly, 
such particle velocities can be effectively measured through experimental methods such as particle tracking velocimetry \cite{Ha15} and ultrasonic correlation spectroscopy \cite{Co00},
and, hence, their corresponding fluctuations can be studied experimentally across
length scales.
In active matter, the velocity-velocity correlation function $C(r)$ for active particles is utilized to define a long-range velocity correlation that arises from motility-induced phase separation (MIPS) \cite{Bu13,Pa13,He20,Ab24}.
In Ref. \cite{He20}, it was shown that $C(r) \sim \exp(-r/\xi)/[\xi^{3/2} r^{1/2}]$ in the MIPS state, meaning that this specific state is typically nonhyperuniform.

In addition, there has been further research on 
hyperuniform states of active matter
that exhibit novel behaviors and physical properties~\cite{lei_nonequilibrium_2019, 
galliano_two-dimensional_2023, kuroda_microscopic_2023, backofen_nonequilibrium_2024, 
zheng_universal_2024, de_luca_hyperuniformity_2024, padhan_suppression_2025}, 
which indicates a rapidly growing area for future research.

 While this article focused on the generalization of the hyperuniformity concept to particle systems in which the particles carry weights (internal degrees of freedom), we note
that another important extension is to that of hyperuniformity of
fields (scalar-valued \cite{To02a,Pe93,Sa03,Wi08,Ma17}, vector-valued \cite{To02a,Ba59,Mo75,Ja90} or tensor-valued \cite{To02a,Ba59,Mo75} fields) that carry weights. Random fields here refer to general random measures, incorporating point configurations and multiphase media as special cases. Such a generalization was already carried out in Ref. \cite{To16a} for scalar fields  and vector fields, 
but not for fields that carry weights.  In Appendix~\ref{app:weigted_field}, we present the relevant equations that generalize hyperuniformity to a special instance of  weighted scalar fields in order to get a sense of the broad applicability of our approach.
Among other things, we identify exact conditions under which a weighted field
inherits hyperuniformity of the original field, whether
hyperuniform or nonhyperuniform not.

\begin{acknowledgments}

S.T., P.J.S., and J.K. were supported by the Army Research Office under Cooperative Agreement Number W911NF-22-2-0103.
J.K. was also supported by the InnoCORE program of the Ministry of Science and ICT (GIST InnoCORE KH0830).
M.A.K.~was supported by the Initiative and Networking Fund of the Helmholtz Association through the Project ``DataMat''
and by the Deutsche Forschungsgemeinschaft (DFG, German Research Foundation) through the SPP 2265, under grant numbers KL 3391/2-2, LO 418/25-1, and ME 1361/16-1.

\end{acknowledgments}

\section*{Data Availability Statement}

The data that supports the findings of this article are proprietary, but
may be available upon reasonable request from the authors.

\appendix

\section{Weight-Averaged Structure Factor Under Periodic Boundary Conditions}
\label{general-S}

To determine the weight-averaged structure factor computationally, we consider 
$N$ weighted points within a fundamental cell occupying region $\Omega$ of volume $V_F$
under periodic boundary conditions. The weight-averaged structure factor  $\weighted{{\cal S}}{\vect{f}}{\vect{k}}$ for a single large configuration
is defined as
\begin{equation}
 \weighted{{\cal S}}{\vect{f}}{\vect{k}}
 \equiv 
 \frac{1}{N} \weighted{\tilde{n}^*}{\vect{f}}{\vect{k}} \Bigcdot \weighted{\tilde{n}}{\vect{f}}{\vect{k}} ,
\end{equation}
where $\weighted{\tilde{n}}{\vect{f}}{\vect{k}}$ is the Fourier transform of (\ref{n}) and, hence,
\begin{equation}
 \weighted{\tilde{n}}{\vect{f}}{\vect{k}} = \sum_{j=1}^N \vect{f}_j e^{-i {\bf k}\cdot {\bf r}_j},
\end{equation}
and $\weighted{\tilde{n}^*}{\vect{f}}{\vect{k}} \equiv [\weighted{\tilde{n}}{\vect{f}}{\vect{k}}]^*.$
\begin{widetext}
\noindent Ensemble averaging $\weighted{{\cal S}}{\vect{f}}{\vect{k}}$ over the configurations with the same particle number and the same fundamental cell yields
\begin{align*}
 \weighted{S}{\vect{f}}{\vect{k}} 
\equiv& 
\frac{1}{N}\E {\sum_{j=1}^N \sum_{l=1}^N \vect{f}_j^* \Bigcdot \vect{f}_l e^{-i {\bf k}\cdot ({\bf r}_j-{\bf r}_l)}} 
=
\frac{1}{N} 
\E{ \sum_{j=1}^N \vect{f}_j^* \Bigcdot \vect{f}_j} 
+ 
\frac{1}{N}  \E{ \sum_{j\neq l}^N \vect{f}_j^* \Bigcdot \vect{f}_l e^{-i {\bf k}\cdot ({\bf r}_j-{\bf r}_l)} }  \nonumber\\
= & 
\E{\abs{\vect{f}}^2}
+ \frac{N(N-1)}{N } \int_\vect{f} \int_\vect{f} \int_\Omega \int_\Omega P_2(\vect{r}_{12};\vect{f}_1,\vect{f}_2) \vect{f}_1^* \Bigcdot \vect{f}_2 e^{-i {\bf k}\cdot {\bf r}_{12}} d \vect{f}_1 d \vect{f}_2 \dd{\bf r}_1 \dd{\bf r}_2 
\\
= & 
\E{\abs{\vect{f}}^2} 
+ 
\frac{N(N-1)}{N/V_F} \int_\vect{f} \int_\vect{f} \int_\Omega P_2(\vect{r};\vect{f}_1,\vect{f}_2) 
\vect{f}_1^* \Bigcdot \vect{f}_2 e^{-i {\bf k}\cdot {\bf r}} d \vect{f}_1 d \vect{f}_2 \dd{\bf r} 
\\
= & 
\E{\abs{\vect{f}}^2} 
+ 
\frac{N(N-1)}{N/V_F} \int_\vect{f} \int_\vect{f} \int_\Omega \qty[ P_2(\vect{r};\vect{f}_1,\vect{f}_2) - \frac{\fn{p}{\vect{f}_1}\fn{p}{\vect{f}_2}}{{V_F}^2}] 
\vect{f}_1^*\Bigcdot \vect{f}_2 e^{-i {\bf k}\cdot {\bf r}} d \vect{f}_1 d \vect{f}_2 \dd{\bf r} 
\\
&+
\frac{N(N-1)}{N/V_F} \int_\vect{f} \int_\vect{f} \int_\Omega \frac{\fn{p}{\vect{f}_1}\fn{p}{\vect{f}_2}}{{V_F}^2} 
\vect{f}_1^*\Bigcdot \vect{f}_2 e^{-i {\bf k}\cdot {\bf r}} d \vect{f}_1 d \vect{f}_2 \dd{\bf r} 
\\
=& 
\E{\abs{\vect{f}}^2} 
+ 
\frac{N(N-1)}{N/V_F} \int_\vect{f} \int_\vect{f} \int_\Omega \qty[ P_2(\vect{r};\vect{f}_1,\vect{f}_2) - \frac{\fn{p}{\vect{f}_1}\fn{p}{\vect{f}_2}}{{V_F}^2}] \vect{f}_1^* \Bigcdot \vect{f}_2 e^{-i {\bf k}\cdot {\bf r}} d \vect{f}_1 d \vect{f}_2 \dd{\bf r} 
+
\frac{(N-1)\E{\vect{f}^*}\Bigcdot\E{\vect{f}}}{ V_F} \int_\Omega e^{-i {\bf k}\cdot {\bf r}} 
\end{align*}

Taking the thermodynamic limit in the last line of the previous equation yields
\begin{align}
 \weighted{S}{\vect{f}}{\vect{k}}
=&
\E{\abs{\vect{f}}^2} + \frac{1}{\rho} \int_\vect{f} \int_\vect{f} \int_\Omega \qty[N(N-1) \fn{P_2}{\vect{r};\vect{f}_1,\vect{f}_2}-\rho^2 \fn{p}{\vect{f}_1}\fn{p}{\vect{f}_2}] \vect{f}_1^* \Bigcdot \vect{f}_2 e^{-i {\bf k}\cdot {\bf r}} d \vect{f}_1 d \vect{f}_2 \dd{\bf r} + (2\pi)^d \rho \E{\abs{\vect{f}}^2} {\tilde\delta}({\bf k}) 
\nonumber \\
= & 
\E{\abs{\vect{f}}^2} + \rho \int_\vect{f} \int_\vect{f} \int_\Omega \fn{h}{\vect{r};\vect{f}_1,\vect{f}_2} \vect{f}_1^* \Bigcdot \vect{f}_2 e^{-i {\bf k}\cdot {\bf r}} d \vect{f}_1 d \vect{f}_2 \dd{\bf r} + (2\pi)^d \rho \E{\abs{\vect{f}}^2} {\tilde\delta}({\bf k}) 
\nonumber \\
=& \weighted{\tilde{\chi}}{\vect{f}}{\vect{k}}/\rho + (2\pi)^d \rho \E{\abs{\vect{f}}^2} {\tilde\delta}({\bf k}),
\end{align}
where we have used relations (\ref{h}) and (\ref{spec-1}). Note that this weight-averaged structure factor is identical
to relation \eqref{eq:Sofk_weighted} for the spectral density, except for the forward-scattering contribution $ (2\pi)^d \rho \E{\abs{\vect{f}}^2} {\tilde\delta}({\bf k})$.
\end{widetext}

\section{Distributions of the Cell Volumes and Excess Side Numbers for 2D models Studied in Secs. \ref{volume} and \ref{charges}.}
\label{prob}

\begin{figure*}[bthp]
 \subfloat[]{\includegraphics[width=0.4\textwidth]{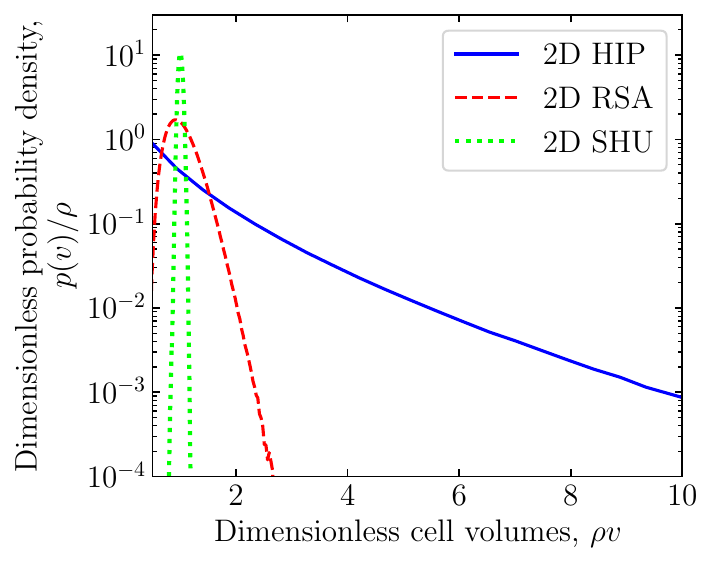}}
 \subfloat[]{\includegraphics[width=0.4\textwidth]{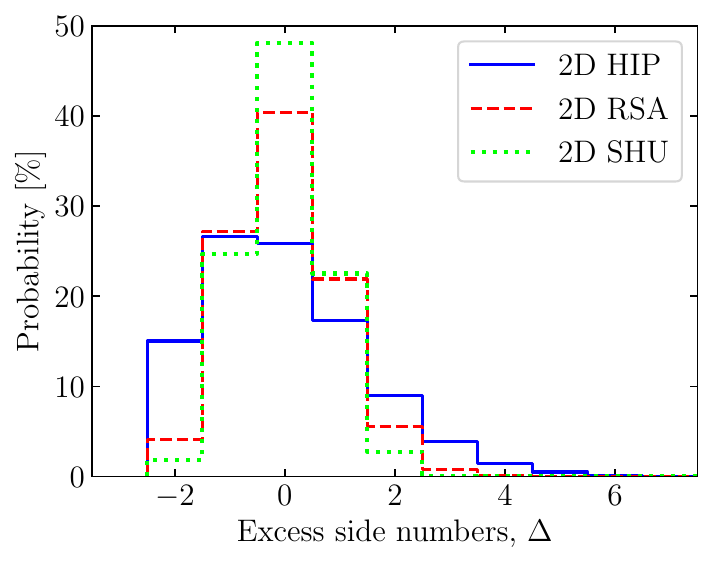}}
 \caption{(a) Dimensionless probability density functions $P(v)/\rho$ of the dimensionless Voronoi-cell volumes $\rho v$ of 2D models of point configurations: HIP, RSA, and SHU. 
 By definition, the mean values of dimensionless Voronoi-cell volumes are equal to unity for all models, i.e., $\expval{\rho v} =1$.
 (b) Probability distributions of the excess side numbers $\Delta$ of the 2D point configurations in (a). 
 For all models, the mean values of the excess side number are equal to zero, i.e., $\expval{\Delta}=0$.
 \label{fig:stat-2D-cells}}
\end{figure*}

Figure \ref{fig:stat-2D-cells} shows the probability density functions
 of the Voronoi-cell-volume weights of the 2D models examined in Sec. \ref{volume} 
as well as probability distributions of the excess-side-number weights of the 2D models
studied in Sec. \ref{charges}. 
We clearly see that HIP is significantly different from other models.
As shown in Fig. \ref{fig:stat-2D-cells}(a), HIP can have an arbitrarily large cell in that the distribution has a power-law tail, reflecting its hyperfluctuating nature.
In contrast, SHU and RSA exhibit compact supports for their Voronoi-cell volumes.
Furthermore, HIP shows a significantly wider range of excess side numbers, denoted as $\Delta$, and possesses an asymmetric distribution; see Fig. \ref{fig:stat-2D-cells}(b). 
This highlights the irregular nature of its Voronoi cells compared to RSA and SHU.
It is noteworthy that every disordered model has a zero mean value for the excess side number, i.e., $\expval{\Delta}=0$. However, this global neutrality alone is insufficient to account for the hyperuniformity observed in the corresponding weighted point configurations. For further details, refer to Fig. \ref{fig:2D-RSA-shuffled} and its accompanying description.

\section{Excess Contributions of the Generalized Spectral Densities}

\begin{figure*}[bthp]
 \subfloat[]{\includegraphics[width=0.3\textwidth]{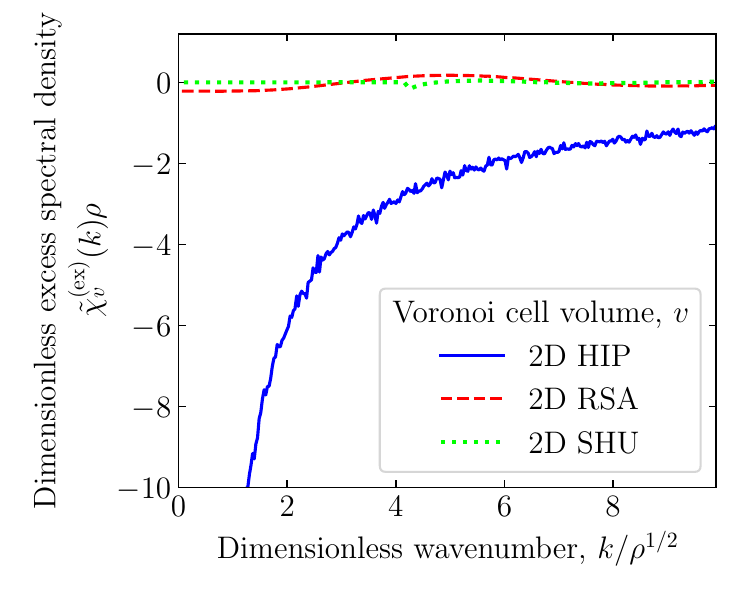}}
 \subfloat[]{\includegraphics[width=0.3\textwidth]{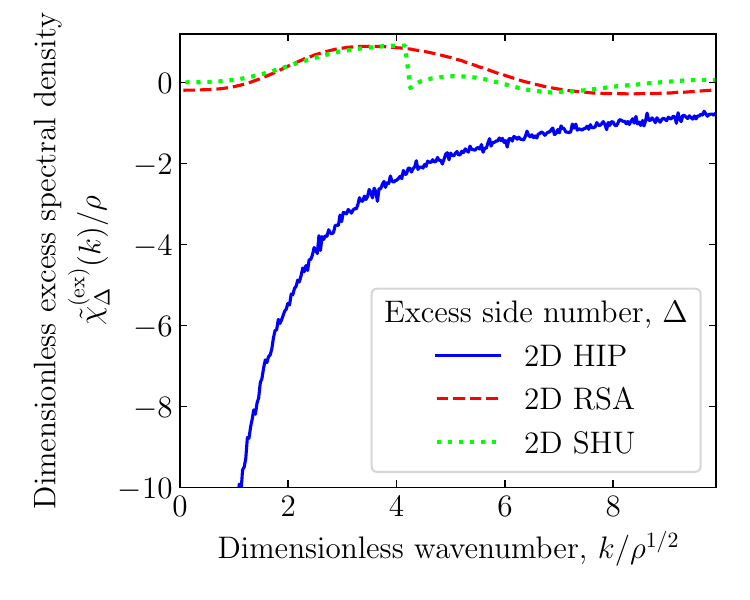}}
 \subfloat[]{
 \includegraphics[width=0.3\textwidth]{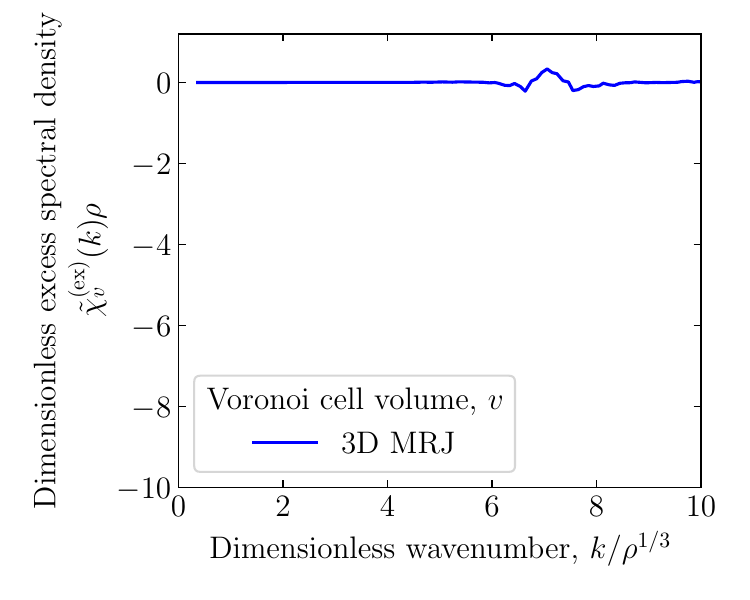}}

 \caption{
 Comparisons of the dimensionless excess spectral densities versus the dimensionless wave number $k/\rho^{1/d}$ for 2D and 3D weighted point
configurations studied in Secs. \ref{volume} and \ref{charges}.
 (a) 2D point configurations weighted by the Voronoi cell area, $v$.
 (b) 2D point configurations weighted by the excess side number, $\Delta$.
 (c) 3D MRJ packings weighted by the Voronoi-cell volume, $v$.
 \label{fig:ex-chif}
 }
\end{figure*}

Here, we show the dimensionless excess spectral density of all the 2D and 3D weighted point configurations examined in Secs. \ref{volume} and \ref{charges}; see Fig. \ref{fig:ex-chif}.
For all 2D models with Voronoi-cell-volume weights, HIP shows the largest amount of excess contributions that are associated with converting antihyperuniform unweighted configurations into hyperuniform weighted ones, and RSA exhibits the next-largest excess contribution. Not surprisingly,
SHU shows the smallest amount of excess contribution, since its unweighted counterpart is already hyperuniform. 
A similar trend is also observed for the 2D models with weights that are the excess side numbers, as shown in Fig. \ref{fig:ex-chif}(b).
The case of 3D MRJ packings with Voronoi-cell volumes makes small but clear excess contributions to the spectral density for the long- and intermediate-wavelength regime ($k/\rho^{1/3} < 10$); see Fig. \ref{fig:ex-chif}(c).
However, the changes for short wavelengths are negligibly small because for an individual particle, the Voronoi-cell volume is nearly identical to unity.

\section{Class I Hyperuniformity Order Metric}
\label{app:Lambda_f}

Because of the slow convergence of the exact integral expression of  the hyperuniformity order metric $\overline{\Lambda}_{\vect{f}}$, defined by Eq. \eqref{eq:Lambda-exact}, it is more efficient to numerically compute it by using the following modified form
in which the upper limit is taken to be a large finite number $Q$:
\begin{align}
\overline{\Lambda}_{\vect{f}}(Q) \equiv &
 \frac{v_{1}(D)}{D} \frac{d}{\pi} \int_{0}^Q \frac{\tilde{\chi}_{\vect{f}}(k)-\tilde{\chi}_{\vect{f}}(0)}{k^2} \dd{k} 
\nonumber \\
=& \overline{\Lambda}_{\vect{f}} - c /Q, \quad Q\to \infty, \label{eq:Lambda-fit}
\end{align}
where $c$ is a positive constant, the latter of which can be viewed as
an ``ultraviolet'' correction term.
By computing the definite integral and fitting with the correction term, one can accurately estimate the value of $\overline{\Lambda}_{\vect{f}}$ from the spectral density for a relatively small range of wave numbers.
The resulting values of the hyperuniformity order metrics of all the weighted models considered here are listed in Table \ref{tab:Lambda}.

\begin{table}[h]
 \caption{Hyperuniformity order metrics of weighted point configurations considered in this work.
 Here, $\overline{\Lambda}_v$ and $\overline{\Lambda}_\Delta$ represent the metrics associated with the Voronoi-cell volume and excess side number, respectively. 
 The values are estimated from Eq. \eqref{eq:Lambda-fit} by taking a characteristic length scale to be $D=\rho^{-1/d}$. Standard errors
 for the fits are given within the parentheses. 
  \label{tab:Lambda}} 
\begin{tabular}{l |c |c }
 \hline 
 Model& $\overline{\Lambda}_v$& $\overline{\Lambda}_\Delta$\\
 \hline 
 1D Poisson& 1.00 (exact)  & - \\ 
 \hline 
 2D HIP& 7.499064(2)& 4.645114(4)\\
 2D RSA& 0.661423(3)& 1.386876(1)\\
 2D SHU& 0.519403(2)& 0.967406(1)\\
 \hline 
 3D MRJ& 0.828619(151)& -  \\
 \hline
\end{tabular}
\end{table}

\section{Hyperuniformity Exponents for 2D Models}
\label{app:exponents}

In this work, the hyperuniformity exponent of a model was estimated by performing a linear regression on the logarithms of wave numbers and $\tilde{\chi}_f(k)$ values for wave numbers up to $f_\mathrm{fit}$.
For 2D weighted models (i.e., HIP, RSA, and SHU) considered in Sec. \ref{s:volumes_2D} and Sec. \ref{charges}, we estimated the hyperuniformity exponents $\alpha_v$ and $\alpha_\Delta$ for several different system sizes.
The estimated exponents for 2D HIP, RSA, and SHU are listed in Tables \ref{tab:hip}, \ref{tab:rsa}, and \ref{tab:shu}, respectively.
We do not tabulate the exponents of the other models here, since we consider a single system size for each model.

\begin{table}[h!]
    \caption{Table of hyperuniformity exponents estimated for the 2D weighted HIP point configurations. 
    The quantities $\alpha_v$ and $\alpha_\Delta$ are the exponents when the particle weights are cell volumes and excess side numbers, respectively.
    Here, $k_\mathrm{fit}$ represents the fitting range. 
    The values in the parentheses are the standard errors. 
    }
    \label{tab:hip}

    \centering
    \begin{tabular}{r|l l |l l}
    \hline 
    $\expval{N}$&   $\alpha_v$& $k_\mathrm{fit}\rho^{-1/2}$ & $\alpha_\Delta$& $k_\mathrm{fit}\rho^{-1/2}$\\
    \hline 
$70^2$&	3.135 (61)&	 0.53&		2.278 (107)&	2.278 (107)\\
$100^2$&	3.148 (63)&	 0.40&		2.362 (30)&	2.362 (30)\\
$200^2$&	3.238 (39)&	 0.26&		2.401 (42)&	2.401 (42)\\
$300^2$&	3.495 (98)&	 0.08&		2.336 (9)&	2.336 (9)\\
$400^2$&	3.132 (18)&	 0.26&		2.287 (19)&	2.287 (19)\\
$600^2$&	3.182 (9)&	 0.33&		2.212 (21)&	2.212 (21)\\
$800^2$&	3.205 (9)&	 0.24&		2.126 (40)&	2.126 (40)\\
\hline
    \end{tabular}
\end{table}

\begin{table}[h!]
    \caption{Table of hyperuniformity exponents estimated for the 2D weighted RSA point configurations. 
    The quantities $\alpha_v$ and $\alpha_\Delta$ are the exponents when the particle weights are cell volumes and excess side numbers, respectively.
    Here, $k_\mathrm{fit}$ represents the fitting range. 
    The values in the parentheses are the standard errors. 
    }
    \label{tab:rsa}

    \centering
    \begin{tabular}{r|l l |l l}
    \hline 
    $N$&   $\alpha_v$& $k_\mathrm{fit}\rho^{-1/2}$ & $\alpha_\Delta$& $k_\mathrm{fit}\rho^{-1/2}$\\
    \hline 
400&	3.941 (10)&	 0.70&	2.656 (10)&	 0.70\\
2000&	3.978 (7)&	 0.58&	2.207 (15)&	 0.31\\
5000&	3.999 (14)&	 0.27&	2.108 (19)&	 0.20\\
10000&	3.918 (5)&	 0.85&	2.086 (23)&	 0.22\\
40000&	4.002 (7)&	 0.51&	2.030 (21)&	 0.20\\
90000&	3.946 (4)&	 0.94&	2.084 (19)&	 0.21\\
\hline
    \end{tabular}
\end{table}

\begin{table}[h!]
    \caption{Table of hyperuniformity exponents estimated for 2D weighted SHU point configurations. 
    The quantities $\alpha_v$ and $\alpha_\Delta$ are the exponents when the particle weights are cell volumes and excess side numbers, respectively.
    Here, $k_\mathrm{fit}$ represents the fitting range. 
    The values in the parentheses are the standard errors. 
    }
    \label{tab:shu}

    \centering
    \begin{tabular}{r|l l |l l}
    \hline 
    $N$&   $\alpha_v$& $k_\mathrm{fit}\rho^{-1/2}$ & $\alpha_\Delta$& $k_\mathrm{fit}\rho^{-1/2}$\\
    \hline 
4000&	3.704 (35)&	 0.25&	2.246 (35)&	 0.25\\
10000&	3.525 (14)&	 0.39&	2.088 (27)&	 0.20\\
20000&	3.530 (33)&	 0.34&	1.977 (83)&	 0.15\\
\hline
    \end{tabular}
\end{table}

\section{Extension to Weighted Random Fields}
\label{app:weigted_field}


The concept of weighted random fields has been employed in various contexts in
physics and engineering. Examples include 
($i$) the definitions of macroscopic fields (e.g., electric displacement and
magnetic field intensity in electromagnetism \cite{Ja90}),
($ii$) Gauge theories from condensed matter to particle physics~\cite{zee_quantum_2003}, 
($iii$) nonstationary (heteroscedastic) Gaussian random fields (with a spatially varying
variance)~\cite{stein_interpolation_1999},
($iv$) random fields on manifolds~\cite{adler_random_2007},
($v$) procedural noise composition in computer graphics~\cite{lagae_survey_2010},
($vi$) visualization of dense vector fields \cite{Sh97b}, and
($vii$) deep learning for image classification, from convolutional neural
networks~\cite{Gu17, An18} to vision transformers~\cite{ViT}.
For simplicity and illustration purposes here, we consider scalar fields
and their weighted counterparts. We begin by first summarizing the key
equations that enable a diagnosis of hyperuniformity of scalar fields,
as generalized by Torquato \cite{To16a}. 
The extension to vector fields is formally the same, which was also
treated in Ref. \cite{To16a}, but it will not be explicitly given here.

\subsection{Brief review of hyperuniformity of scalar fields}

Following Ref. \cite{To16a}, consider a statistically homogeneous random real-valued scalar field $F(\bf x)$ in $\mathbb{R}^d$ with an autocovariance function
\begin{equation}
\chi_{_F}({\bf r})= \expval{    [F({\bf x}_1)- \langle F({\bf x}_1)\rangle]\, [F({\bf x}_2) -  \langle F({\bf x}_2)\rangle] \,},
\label{spec-field}
\end{equation}
where we have invoked the statistical homogeneity of the field, meaning that the autocovariance function depends solely on ${\bf r}={\bf x}_2 -{\bf x}_1$. We assume that 
the associated spectral density ${\tilde \chi}_{_F}({\bf k})$ exists. 
The hyperuniformity condition
is simply that the non-negative spectral density obeys the 
small-wave-number condition:
\begin{equation}
\lim_{|{\bf k}| \rightarrow {\bf 0}} {\tilde \chi}_{_F}({\bf k})=0,
\label{hyp-field}
\end{equation}
which is equivalent to the direct-space sum rule
\begin{equation}
\int_{\mathbb{R}^d} \chi_{_F}({\bf r}) \dd{\bf r}=0.
\end{equation}

The local variance associated with fluctuations in the field, denoted by $\sigma_{_F}^2(R)$, 
is related to the autocovariance function or spectral function  as follows:
\begin{eqnarray}
\sigma^2_{_F}(R)&=&  \int_{\mathbb{R}^d} \chi_{_F}(\mathbf{r}) \alpha_2(r; R) \dd{\mathbf{r}},\nonumber \\
&=&\frac{1}{ (2\pi)^d} \int_{\mathbb{R}^d} {\tilde \chi}_{_F}({\bf k})
{\tilde \alpha}_2(k;R) \dd{\bf k},
\label{local-scalar}
\end{eqnarray}
where $ \alpha_2(r; R)$ and  ${\tilde \alpha}_2(k;R)$ are given in Eqs. \eqref{eq:alpha_direct} and \eqref{eq:alpha-fourier}, respectively.
Note that the definition of the 
local variance used in Ref. \cite{To16a} differs from Eq. \eqref{local-scalar} by a multiplicative factor of $1/v_1(R)$.

\subsection{Generalization to weighted random fields}\vspace{-0.1in}

We now consider a  weighted generalization of a statistically homogeneous random scalar field $F(\bf x)$ in $\mathbb{R}^d$.
One possible way to weight the field  $F({\bf x})$ is via a convolution
operator, i.e.,   convolution of $F(\bf x)$ with a kernel $K({\bf x};{\bf C})$
to produce a ``weighted'' field $W({\bf x})$:
\begin{align}
W({\bf x}) = \int_{\mathbb{R}^d} F({\bf x}^\prime) K({\bf x}-{\bf x}^\prime;{\bf C})  \dd{\bf x}^\prime,
\label{field}
\end{align}
where $K({\bf x};{\bf C})$ is a non-negative dimensionless scalar  kernel function that is a function of position $\bf x$ and sufficiently localized so that its Fourier transform exists. Here $\bf C$ represents a set
of parameters that characterize the kernel.
It is seen that the kernel locally smooths out the original field.
A simple example is the Gaussian kernel function
\begin{equation}
K({\bf r};a) =\exp(-(r/a)^2),
\end{equation}
where $a$ is a characteristic length scale that is proportional to the standard
deviation of the Gaussian. The corresponding Fourier transform  is given by
\begin{equation}
{\tilde K}({\bf k};a) =\pi^{d/2} a^d\exp[-(ka)^2/4].
\end{equation}

For any choice of a Kernel $K(\vect{x};\vect{C})$, the definition of the weighted field, given in Eq. \eqref{field}, implies that its spectral density can be written as 
\begin{equation}
{\tilde \chi}_W({\bf k})={\tilde K}^2({\bf k}; \vect{C})  {\tilde \chi}_{_F}({\bf k}),
\end{equation}
where ${\tilde \chi}_{_F}({\bf k})$ is the spectral density of the original field defined above.
Thus, the local variance associated with fluctuations in the weighted field $W({\bf x})$ 
is related to the spectral function ${\tilde \chi}_W({\bf k})$ as follows:
\begin{eqnarray}
\sigma^2_{_W}(R)&=& \frac{1}{ (2\pi)^d} \int_{\mathbb{R}^d} {\tilde \chi}_W({\bf k}) {\tilde \alpha}_2(k;R) \dd{\bf k},\nonumber \\
&=&\frac{1}{ (2\pi)^d} \int_{\mathbb{R}^d}  {\tilde K}^2({\bf k};\vect{C})  {\tilde \chi}_{_F}({\bf k}){\tilde \alpha}_2(k;R) \dd{\bf k}.
\label{local-scalar-2}
\end{eqnarray}
We see that the integrand of Eq. \eqref{local-scalar-2}, compared to the one for the original local variance (\ref{local-scalar}), differs exactly
by a factor ${\tilde K}^2({\bf k}; \vect{C})$.
It follows that the direct-space representation of the local variance is given in terms of the autocovariance function ${\chi}_W({\bf r})$ of $W(\vect{x})$:
\begin{eqnarray}
\sigma^2_{_W}(R)&=&  \int_{\mathbb{R}^d} \chi_W({\bf r}) \alpha_2(r;R) \dd{\bf r}.
\label{local-scalar-3}
\end{eqnarray}
Physical examples of such weighted continuous fields are experimental
measurements with point-spread functions~\cite{goodman_introduction_2005}, diffusion in random
potentials~\cite{marconi_dynamic_1999, archer_dynamical_2004, te_vrugt_classical_2020}, 
and image processing of random fields in physics (like
the cosmic microwave background)~\cite{starck_sparse_2010, marinucci_random_2011}.
This exemplary use of a kernel to obtain a weighted field can be further
generalized, e.g., using random correlated kernels as in the recent
preprint in mathematics \cite{Kl25b}.
\vspace{0.05in}
 
{\bf Remarks}:
\vspace{-0.1in}

\begin{itemize}
\item From relation (\ref{local-scalar-2}), we see that if the original field is hyperuniform,
then the variance of the weighted field will grow more slowly than $R^d$, even if ${\tilde K}({\bf 0};{\bf C}) \neq 0$,
implying a hyperuniform weighted field.
\item However, we observe that if the original field is nonhyperuniform, then the weighted
field can be hyperuniform only if the product ${\tilde K}^2({\bf k};\vect{C})  {\tilde \chi}_{_F}({\bf k})$
tends to zero in the limit $|\bf k| \to 0$.
\end{itemize}

\end{document}